\documentclass{article}
\usepackage[utf8]{inputenc}
\usepackage{xcolor}
\usepackage{authblk}
\usepackage[caption=false]{subfig}

\title{Diversity of Exoplanets} 
\author[1,2,3,4]{Diana Valencia}
\author[5,6]{Amaya Moro-Martin}
\author[7]{Johanna Teske}
\affil[1]{Department of Astronomy, University of Toronto, Toronto, ON, M5S 3H4, Canada}
\affil[2]{Department of Physical Sciences, University of Toronto Scarborough, Toronto, ON, M5S 3H4, Canada}
\affil[3]{Physics Department, University of Toronto, Toronto, ON, M1C 1A4, Canada}
\affil[4]{Observatoire de la Cote d'Azur, Boulevard de l'Observatoire, Nice, 06300, France}
\affil[5]{Space Telescope Science Institute, 3700 San Mart\'in Dr., Baltimore, MD 21218, USA}
\affil[6]{William H. Miller III Department of Physics and Astronomy, Johns Hopkins University, 3400 N. Charles Street, Baltimore, MD 21218, USA}
\affil[7]{Carnegie Science Earth and Planets Laboratory, 5241 Broad Branch Road, NW, Washington, DC 20015, USA}

\date{}
\usepackage{natbib}
\usepackage{graphicx}
\usepackage{geometry}
\usepackage[utf8]{inputenc}
\usepackage{amssymb}

\begin{document}
\maketitle
\begin{abstract}

This review article delves into the study of low-mass exoplanets: super-Earths, mini-Neptunes and the new categories within and between that we are starting to discover. We provide an overview of current exoplanet observational capabilities, their limitations, and what they are allowing us to learn about low-mass planets. We briefly summarize the most important aspects of planet formation, with an emphasis on processes that may be testable with small exoplanets, in particular those that affect their composition. 
We also describe the observed compositional diversity of low-mass exoplanets and what it teaches us about planet formation pathways. We finish this review summarizing the study of the composition of small exoplanets during the very last stage of stellar evolution, by studying white dwarfs. This review is written as the JWST is making its first contributions to small planet characterization, rapidly opening new lines of inquiry. 
   
\end{abstract}
\color{teal}
\section{Key points}
\begin{itemize}
    \item The formation of planets starts from the condensation of the first solids from the protoplanetary gas disk, the growth to pebbles, planetesimals, embryos and eventually full fledged planets, while experiencing a variety of processes such as collisions, migration, atmospheric evaporation, etc. The composition of the planets reflects this complex history, and the hope is that exoplanets can help us parse it together.
    \item Rocky exoplanets can help determine whether a primordial composition is a natural formation outcome by comparing the refractory relative abundances of planets to that of their host stars. While rocky planets seem to have a larger compositional variety compared to that of stars, the data is insufficient yet to establish whether there is a correlation.     
    \item Determining whether low-mass planets have atmospheres with space telescopes like JWST, and for those that do, their composition, is a crucial step to understanding the solar system. The ultrashort period planets are a unique opportunity to study magma-ocean/atmosphere thermochemical exchange relevant to the origin of the Earth. 
    \item As the star approaches the end of its lifetime and becomes a white dwarf, planets and planetesimals may be scattered into its tidal radius and become tidally disrupted, offering an exciting and diametrical opportunity to learn about their mineralogy by studying the resulting debris, and the elemental abundances by studying the white dwarf atmospheric pollution. So far, the predominant material observed has been Si, Mg, Fe, O, resembling the main components of the solar system.
\end{itemize}
\color{black}

\section{Introduction} 
\label{Intro}

The existence of planets around other stars had been a centuries old problem until 1995 when the first extra solar planet around a main sequence star was discovered \citep{MayorQueloz95}. 51~Pegasi~b is at least 
half of Jupiter mass and orbits with a period of $\sim$4 days. It produces a signal (the slight movement of the star due to the gravitational tug of the planet) that at the time was just at the cusp of what could be observed. Thanks to technological and analysis improvements, as of spring 2024 there are over 5600 confirmed\footnote{Confirmed planets refer to planet detections where false positives have been ruled out (for example the presence of a stellar companion causing the star's radial velocity signal), or that have been detected using two different techniques (for example the transit and radial velocity methods). If the signal is detected using the same technique but by different telescopes the planet detection is said to be "validated".} exoplanets in about 4160 systems. Most of these planets are large, due to the biases in observations: it is easier to detect planets that have short periods, and have either a large mass or radius -- like 51 Pegasi b. However, low-mass/small exoplanets are now routinely being discovered, thanks to the increased capabilities of space and ground-based telescopes, and the fact that low-mass/small planets are intrinsically more numerous. This review focuses on this planet population, encompassing Earth's, super-Earths, water worlds, and mini-Neptunes.

\subsection{Exoplanet Observations}
\label{ExoplanetObservations}

The main observables in exoplanets are the planet radius and mass, the composition of the atmosphere, the temperature map along meridians, the orbital distance from the star, and the stellar age. 

\subsubsection*{Planet Radius}
The radius is measured for planets in a configuration where they cross in front of their host star periodically, dimming its light with respect to the observer. This method is called the transit method and the measured reduction of flux in the light curve is proportional to the cross-sectional area of the planet with respect to that of the star. Thus, the radius of the planet is known after determining the stellar radius. Due to the randomness of the orbits' orientation in the sky with respect to us, a small proportion of systems are expected to transit ($\sim10\%$ for short period planets, and less for longer period planets based on geometric configurations). The smaller the planet or larger the star, the more difficult it is to measure the planetary radius. This explains why the first planets that were seen in transit were the hot-Jupiters 51~Pegasi~b \citep{Henry:2000} and HD~209458~b \citep{Charbonneau:2000}. Along the same line, it is easiest to discover and follow-up small planets if they orbit M dwarf\footnote{M dwarfs are the smallest type of stars with ongoing thermonuclear fusion. They have temperatures ranging from 2000 K to 3900 K and stellar masses ranging from 0.07 M$_{Sun}$ to 0.6 M$_{Sun}$. They are the most common type of star in the Milky Way.} stars. In general, more massive planets tend to be bigger, but there is considerable spread due to differences in composition that will be discussed in this review. For an in-depth discussion of the transit method see \citet{DeegAlonso:ReviewTransit:2017}.

\subsubsection*{Planet Mass}
The planet mass is normally estimated via the radial velocity method that measures the star's periodic velocity around the system's center of mass when planets are present \citep{MayorQueloz95, Butler:1997}. The orbit of the star is more pronounced and easier to detect when planets with larger masses are exerting a stronger gravitational pull on their host star. 
Owing to the inclination\footnote{The orbital inclination measures the tilt of the planet's orbit with respect to the planet perpendicular to the line-of-sight from Earth} of the system with respect to the observer, what is measured is the minimum mass (e.g. $M\sin i$, where $M$ and $i$ are the mass, and the inclination, respectively). The signal's velocity amplitude scales directly with the planet-to-star mass ratio and inversely with the planet's semi-major axis. Thus, observations are biased towards massive short-period planets. Not surprisingly, the first exoplanet, 55 Pegasi b, was a massive planet orbiting very close to the star. Also, similarly to the transit method, the mass of the planet is known only after factoring for the mass of the star. This highlights the fact that the precision on planetary mass and radius is intrinsically associated to the precision of stellar parameters. For a review on the radial velocity method see \citet{LovisFischer:ReviewRV:2010}.

A second method to measure the planet mass is the transit timing variations (TTV) and transit timing durations (TTD). The timing and duration of the transit of a planet are affected by the gravitational influence of any other planets present in the system. These interactions get amplified when planets are near resonances, such that it is possible to constrain the masses of the planets in the system. The first example of TTVs and TTDs was the Kepler-9 system with two exo-Saturns \citep{Holman:Kepler9:2010}. More recently, one of the best examples of TTVs is arguably that of the Trappist-1 system, with seven planets with masses between $ 0.8-1.2 M_E$ near resonances in a very compact orbital configuration (all planets are within 0.1 AU). Their TTVs allowed for exquisite mass characterisation with errors of only 3-5\% for these low mass planets \citep{Agol:2021}. For a discussion on these and other discovery methods see \citet{Fischer:DetectionTechniques:2014}.

\subsubsection*{Atmospheric Composition from Low-resolution Transmission and Reflected Spectra}

\begin{figure}[htp]
\begin{center}
\subfloat{%
     \includegraphics[scale=2.356]
     {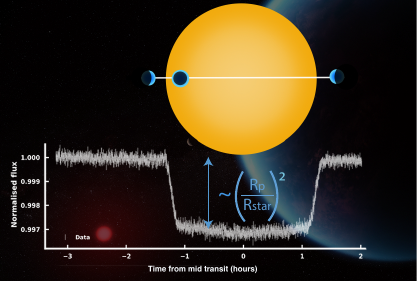}%
     }
     
\subfloat{%
     \includegraphics[scale=0.29]
     {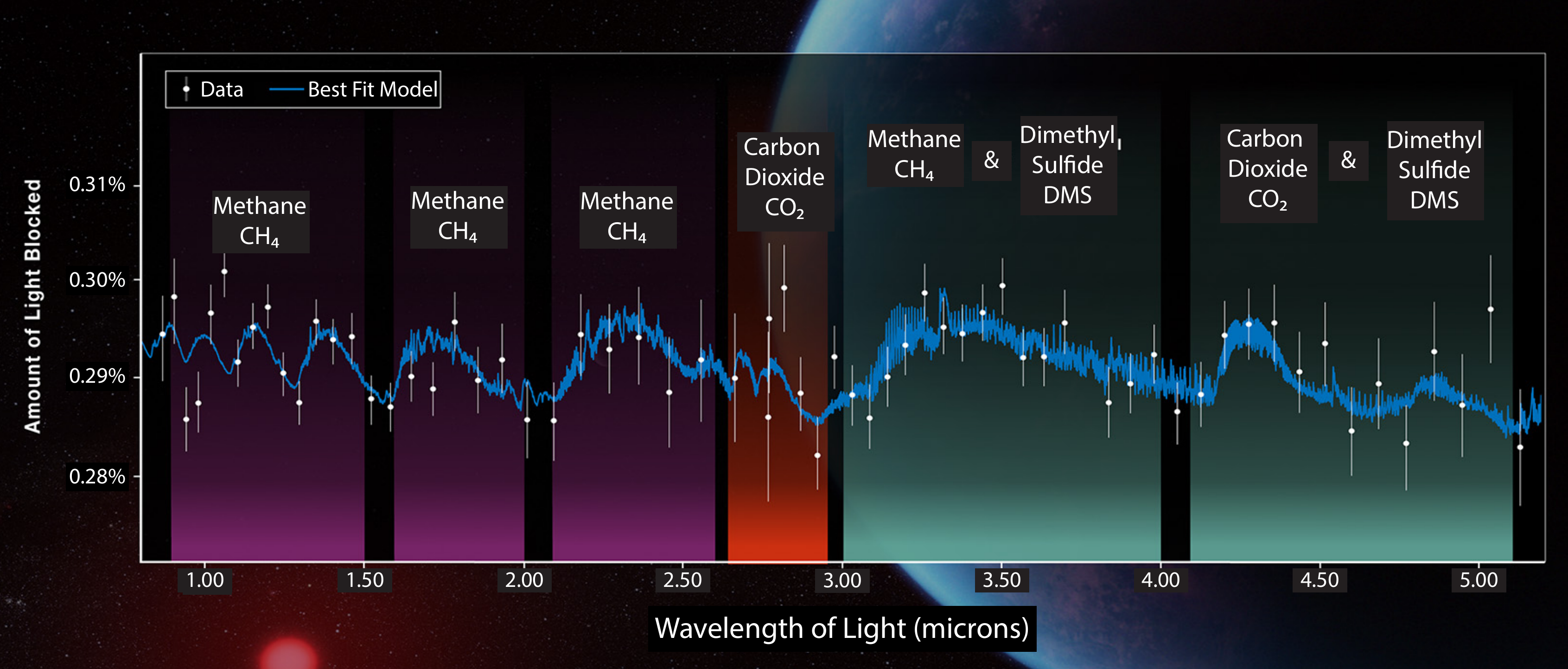}%
     }
     \caption{{\bf Top Figure:} Primary transit of K2-18b obtained with the NIRspec (Near-Infrared Spectrograph) instrument on JWST (light curve from \citet{Madhu23}). {\bf Bottom Figure:} Transmission spectrum of K2-18 b obtained JWST instruments NIRISS (Near-Infrared Imager and Slitless Spectrograph) and NIRSpec. It shows methane, carbon dioxide, and a possible detection of dimethyl sulfide (DMS) consistent with K2-18b being a Hycean world (possessing a liquid water ocean underneath a hydrogen-rich atmosphere). The mass, radius and equilibrium temperature of H2-18b are estimated to be $\sim$ 8.6 M$_E$, 2.61 R$_E$, and $\sim$250-300 K, respectively, for an albedo between 0-0.3 \citep{Benneke:K2-18b}. Illustration credits: NASA, ESA, CSA, Ralf Crawford (STScI), Joseph Olmsted (STScI). Science credits: Nikku Madhusudhan (IoA).}
\label{fig:K2-18b}
\end{center}
\end{figure}

Around bright enough stars, it becomes feasible to probe the atmospheres of transiting planets, if the atmospheres have large enough scale heights\footnote{The scale height is a distance over which the atmospheric pressure decreases by a factor of {\it e}. Large scale heights refer to ``puffy" planets.}. When the planet passes in front of the star, the light from the star that interacts with the terminator\footnote{The terminator is the line between the illuminated and dark side of the planet.} regions of the planet's atmosphere acquires an imprint as seen in the combined stellar+planet spectrum. By removing the stellar component -- when the planet is not crossing the surface of the star along our line-of-sight -- the transmission spectrum of the planet is recovered. This technique is referred to as transmission spectroscopy. It was first theoretically proposed by \citet{Seager:2000}, and first applied in observations of hot-Jupiter HD 209458 b \citep{Charbonneau:2002}. The features in the transmission spectrum are unique to the composition of the atmosphere. This method probes the highest portions of the planet's atmosphere (at the $\sim$ millibar level), where light can still pass through and reach the observer. This direct measure of the planet's spectrum is typically obtained at low resolution (with resolving power\footnote{The resolving power is given by $R = \lambda/\Delta\lambda$, where $\lambda$ is the observing wavelength and $\Delta\lambda$ is the spectral resolution, a measure of the ability to resolve spectral features.} of R $\sim 100-1000$). Interpreting this type of observation is challenging because different compounds produce a similar signal when averaged to low resolution at a particular wavelength range. A solution is to observe at a wide range of wavelengths because of the ability to identify compounds using multiple spectral features; this is one of the main advantages of using the broad bandwidth of JWST spectral modes. See Fig. \ref{fig:K2-18b} for a prime example of this technique on K2-18b. 

\begin{figure}[htp]
\begin{center}
\subfloat{%
     \includegraphics[scale=0.302]
     {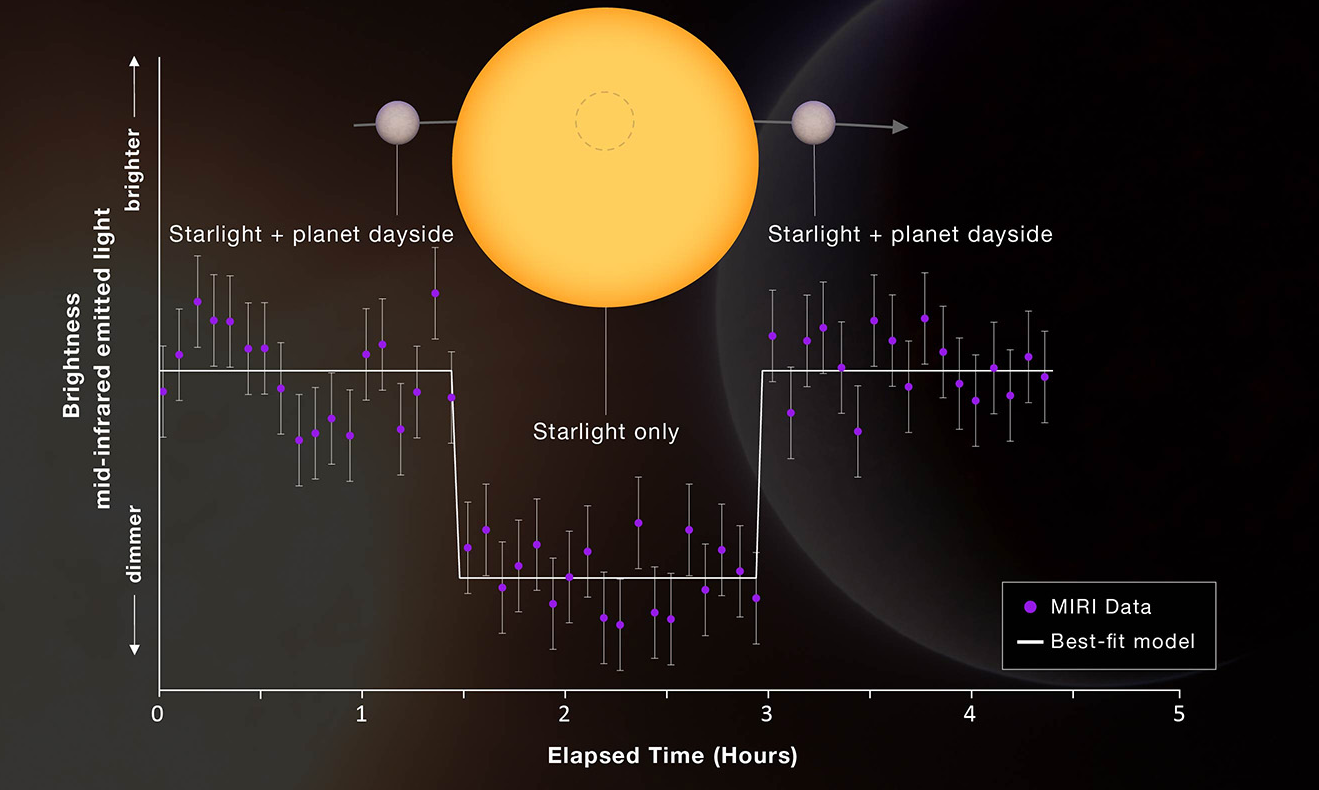}%
     }
     
\subfloat{%
     \includegraphics[scale=0.12]
     {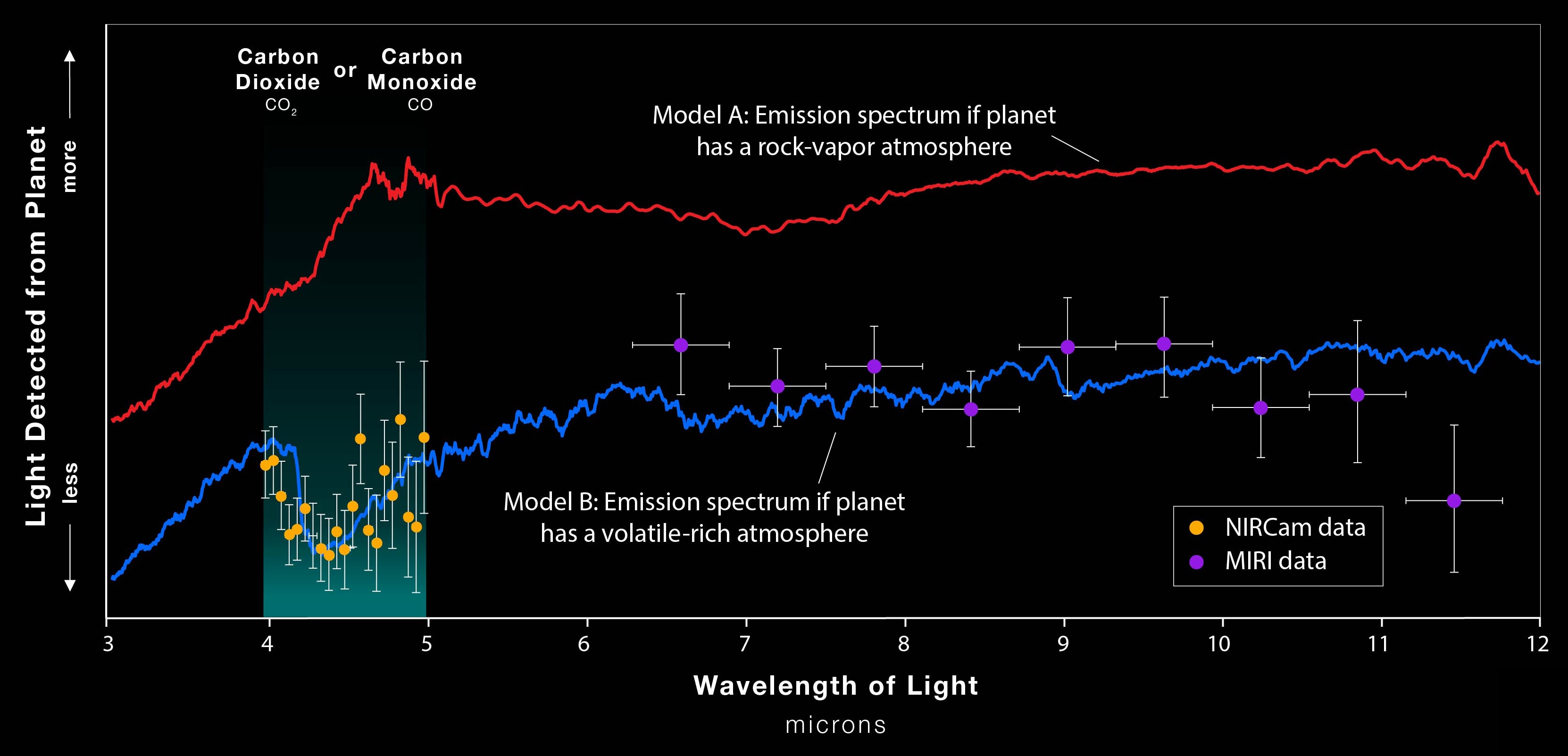}%
     }
     \caption{ {\bf Top Figure:} 55 Cnc-e secondary eclipse taken with MIRI camera on JWST. Its mass and radius are close to $8M_E$ and $1.9 R_E$. The eclipse depth yields a brightness temperature of $1800$, which is lower than the temperature corresponding to the case of zero heat redistribution, indicating the presence of an atmosphere. {\bf Bottom Figure:} Thermal Emission Spectrum of 55 Cnc-e taken with NIRCAM on JWST consistent with an atmosphere composed of CO-CO$_2$ \citep{Hu:2024}.    
     Top figure: Illustration credits: NASA, ESA, CSA, Joseph Olmsted (STScI). Science credits: Aaron Bello-Arufe (NASA-JPL). Bottom figure: Illustration credits: NASA, ESA, CSA, Joseph Olmsted (STScI). Science credits: Renyu Hu (NASA-JPL), Aaron Bello-Arufe (NASA-JPL), Michael Zhang (University of Chicago), Mantas Zilinskas (SRON). 
       }
\label{fig:55Cnc-e}
\end{center}
\end{figure}

As the planet orbits the star, the flux received comes from both the star and planet, either because the planet is reflecting light from the star back (depending on the albedo), or because the planet is emitting its own thermal light (in the infrared). By removing the contribution of the star, obtained when the planet is behind, it is possible to recover the contribution of the planet only. This technique is known as reflectance or emission spectroscopy. This flux is a measure of the temperature of the planet and its albedo. See Fig. \ref{fig:55Cnc-e} for a prime example of this technique on 55 Cnc-e. So far, about 100 planets have transmission spectra taken, 25 only emission spectra, and about 45 have both\footnote{Numbers compiled from the atmospheric spectroscopy list from https://exoplanetarchive.ipac.caltech.edu}. 

Observing the flux of the system as the planet orbits its star results in a phase curve that tracks temperature as a function of the planetary longitude, i.e. a temperature map. If a planet has no atmosphere and is tidally locked (meaning it has a permanent day and night side)\footnote{As the planet orbits near the star, the gravitational gradient across the planet creates a bulge. Because the planet is rotating and there is resistance to reshaping, this bulge has a lag with respect to the planet-star axis. This misalignment creates a torque that with time synchronizes the planet's rotation with its orbital period. It is assumed that many short-period planets are tidally locked because the timescale for tidal locking is smaller than the stellar age, assumed to be an estimate of the planet's age.}, then there is a high contrast between the moment the planet passes in front of the star, when its non-illuminated side is facing the observer, and the moment right before it passes behind the star when the illuminated side is facing the observer. Many short period planets are commonly assumed to be tidally locked, showing the same side to the star during their orbit. If instead the planet has a thick atmosphere that redistributes heats efficiently, the hotspot blurs or vanishes and there is little difference as the planet orbits its star. For non-tidally locked planets, the interpretation of the temperature map is more complex as it depends on the radiative timescale, the planetary rotation relative to synchronous rotation and thermal response timescale of the ground. In either case, monitoring the flux from the planet procures information on whether there is an atmosphere or not, and how effective it is at reflecting heat back and/or redistributing its heat around the planet. For more information on phase curves see \citet{Parmentier:PhaseCurves:2018}.

\subsubsection*{Atmospheric Composition from High Resolution Spectroscopy of the Star and Planet System}

There is another way by which the composition of the atmosphere can be obtained. When observing the system as a whole, the contribution to the total spectrum from the planet is red shifted and blue shifted at higher values than the contribution of the star, given that the planet is moving much faster, at larger semi-major axis around the common the center of mass. This property is exploited with ground observations using high resolution spectroscopy ($R\gtrsim 100000$) that are able to separate each contribution, and observations can capture hundreds to thousands more spectral lines than in low-resolution transmission spectroscopy \citep{Snellen:2013}. These spectral lines are cross-correlated to well known templates of target compounds (e.g. CO, H$_2$O, etc). If there is a strong cross-correlation signal at the frequencies where the blue and red shift from the planet are expected, it is deemed that the target compound is present in the atmosphere of the planet. This technique can be used even if the planet is not transiting. About 40 planetary atmospheres have been characterized with this technique so far, mostly gas giants.

It is important to note though, that with either low-resolution or high-resolution spectroscopy, in emission or transmission, it is not possible to obtain the total abundance of an atmospheric compound. What is measured are the relative abundances. Phase curves, however, may help put some constraints on the amount of mass in the atmosphere that redistributes the heat that is observed. In the case of transmission spectroscopy, it is also true that hazes and clouds dampen considerably features that would otherwise be present in a clear atmosphere as the light is absorbed at all frequencies. It is believed that clouds/hazes are responsible for the featureless spectra of several small exoplanets [e.g. \cite{Kreidberg2014,Dymont2022,Kempton2023}]. In contrast, high-resolution spectroscopy is sensitive to regions higher than where clouds are expected, and so can still retrieve information on the composition of the atmosphere \citep{Hood2020,Gandhi2020}.

As mentioned above, we only have information on detailed atmospheric composition of a few giant planets and couple of low-mass exoplanets. This situation is expected to change rapidly in the era of JWST. Compact planets are more challenging to observe because of their smaller cross sectional area. The interpretation of their atmospheric observations is also more challenging because of interior and atmospheric coupled processes. Given these challenges, the first order of business is to know which compact planets have atmospheres and which ones do not and, if possible, as a second step determine the composition for those that have atmospheres. The first examples of searches for small planet atmospheres show that some very hot super-Earths are likely bare rocks (in the sense that they do not have atmospheres affecting their radius, like the Earth and Venus) \citep{Kriedberg19,Crossfield2022,Greene2023,Ih2023,Zhang2024}, while recent new exciting results with JWST, shown in Fig.\ref{fig:55Cnc-e}, show that 55 Cnc-e has an atmosphere \citep{Hu:2024}.

It is only through the increased capabilities of JWST in terms of wavelength range, spectral resolution and throughput\footnote{Throughput refers to the efficiency to convert the flux of photons received by the telescope's primary mirror to an electron signal by the different instruments at a given wavelength range or filter.} that we are entering an era where it is possible to retrieve the relative abundances of certain molecular compounds for a handful of small planets. These molecules include H$_2$O, CO, CO$_2$, and CH$_4$, but also possibly NH$_3$, SO$_2$, CS$_2$, with features in the near-to-mid-infrared where JWST operates.  
In particular, JWST has opened a unique opportunity to use transmission/emission spectroscopy to study rocky planets around M dwarfs to identify the prevalence and diversity of their atmosphere and even attempt to identify biosignatures. The interest is that not only M dwarfs are the most prevalent stars in the Galaxy, but they also have a higher probability of hosting small planets, so overall M stars are expected to host most rocky planets. In addition, being smaller, dimmer and cooler than the sun, planet transits will be deeper and therefore easier to detect. This is just the beginning of a field that is poised for discoveries.

While the future is promising, it is important to keep in mind the limitations. A main source of noise for obtaining reliable planetary data is the stellar activity. Variable processes observed in the outer stellar atmospheres mainly due to the presence of active magnetic fields, especially in low-mass stars, can be a difficult source of error to correct for. In terms of interpretation of the data, a salient difficulty arises because the data for a particular planet is scarce while linking composition, either atmospheric or bulk, to internal or formation processes, is in general a degenerate endeavour. Because of this, for a single planet it will most likely be very difficult to parse together its formation and evolution history. However, the power of studying exoplanets stems from the large numbers and the hope is to uncover general trends.

\subsection{Nature Produces More Smaller Planets}
\label{Prevalence}

The first planets discovered were all giant planets, with a preference for close-in planets, but this was due to biases in the detection methods. Major discoveries of small planets came after the \emph{Kepler} space mission was launched. Its goal was to look for Earth-like planets (in terms of mass, radius and distance from the star) in a portion of the Milky Way galaxy by continuously staring at about 150,000 stars to detect transits \citep{Borucki2010, Batalha2014, Lissauer2023}. With \emph{Kepler} observations it became clear that nature produces a large diversity of planets. 

\begin{figure}
     \centering     \includegraphics{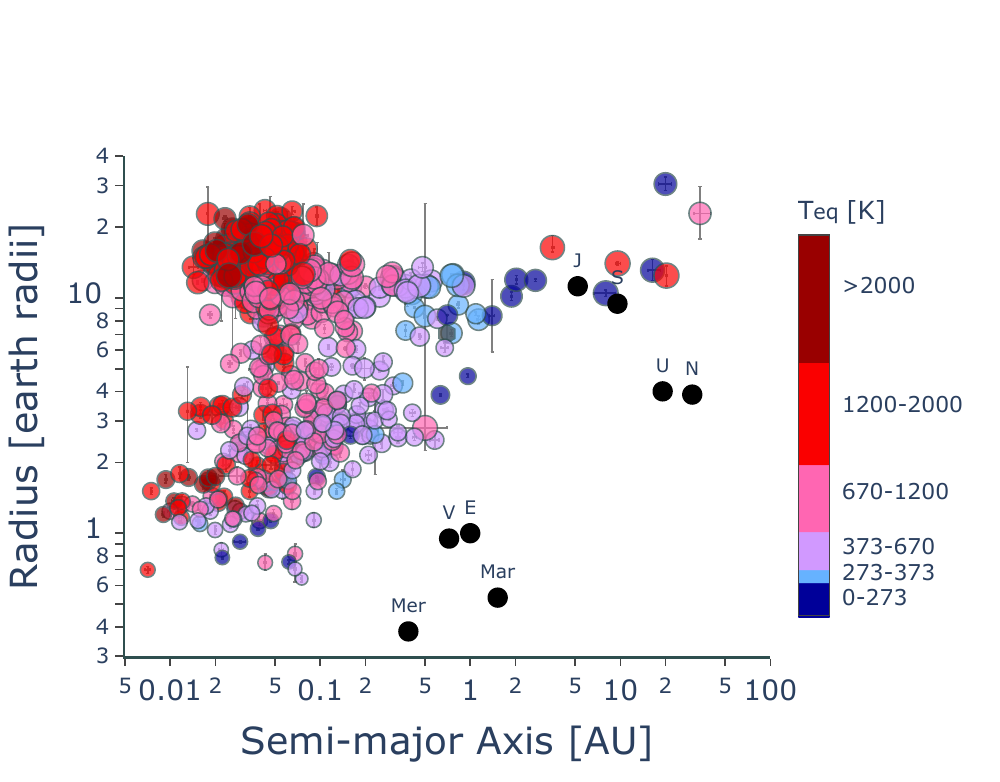}
     \caption{Exoplanets with masses error and radii error less than 25\%. Symbol size is proportional to mass. Colors correspond to equilibrium temperatures calculated with bond albedos of 0.3 and redistribution factors of 4. Colors are loosely binned according to physical processes: planets with icy surfaces should they have any water (0-273 K), planets that allow for liquid water (273-373 K), planets that are in runaway greenhouse below the critical point of water (373-670K), hot planets that cannot have water clouds (670-1200K) and magnetic effects on their atmospheres are negligible, hot planets where magnetic effects like ohmic dissipation are important, very hot planets where magnetic effects are less important (due to full ionization and freezing to magnetic field).
     Solar system planets are labeled and shown as black symbols for reference.}
     \label{fig:aR}
 \end{figure}

Figure \ref{fig:aR} shows all confirmed planets to date (Spring 2024, https://exoplanetarchive.ipac.caltech.edu/) as a function of semi-major axis. This sample is divided into different classes of planets. Radius is a good parameter to broadly distinguish among the different types of planets, given that variations in radius require large variations in composition. Three main categories were originally identified, the exo-Giants including exo-Jupiters and exo-Saturns (close in mass or radius to that of Jupiter or Saturn), the mini-Neptunes that were deemed large enough to require a volatile atmosphere and had sizes close to Neptune and below, and the super-Earths, which were considered compact planets with masses in the range of $1-10M_E$, where the subscript $E$ denotes Earth. As advancements have allowed astronomers to discover even smaller planets than Earth, there is now a category of sub-Earths. These categories were loosely assigned based on mass or radius. However, as it became more obvious that composition was an important discriminant, and that there is an overlap in composition within the $1-10M_E$ and $1-2R_E$ ranges, other terminology started appearing including Water/Icy Worlds, Steam Planets, Hycean Worlds, Super-Mercuries, Super-Moons, etc. In short, the terminology used in exoplanets is dynamic and reflects our need to adapt to new discoveries. For the purpose of this review, low-mass exoplanets comprise all planets that are mostly rocky, have condensate volatiles (water/ice), or have a small volatile envelope, regardless of their period. This seems to coincide with the $\lesssim10M_E$ and $\lesssim2R_E$ mass and radius limits. 

Within the confirmed planets, there appears to be an overabundance of planets of Jupiter-size with periods of 1-10 days around Sun-like stars, the so called Hot Jupiters. As explained above, this is due to biases in detection. Removing these biases reveals that hot Jupiters make only $\sim 1\%$ of the exoplanet population \citep{Wright12}. On the other hand, despite the difficulties in observing smaller planets ($R < 4 R_E$), regular detections of them is indicative of their prevalence. Addressing biases in observations shows that super-Earths and mini-Neptunes \citep{Fressin13,
Petigura13, Mulders15}, and perhaps even sub-Earths \citep{YansongWu21} are the most abundant type of planets. 

In this review, we focus on the low-mass exoplanets for several reasons: 1) They are the most common planets in the Galaxy; 2) They present a large variety of characteristics that can help us test physical processes in a wide parameter space (like testing atmospheric circulation on a rocky planet driven by extreme temperature differences). 3) They are complex planets where interior and atmospheric coupled processes are important, which opens a new window to understand terrestrial planets. Thus, these planets are test beds to understanding Earth and other solar system planets. However, caution needs to be taken as to the type of questions that astronomical observations are able to answer. Owing to their small atmospheres, super-Earths and even mini-Neptunes are predominantly governed by a rocky, and perhaps icy interior, where geophysical and geochemical aspects cannot be ignored. The methodology of combining complex geophysical/chemical aspects under an astrophysical context has slowly become the norm in the study of these planets. However, given the limited scope (and sometimes precision of) astronomical data, there are many degeneracies and care has to be taken to discern which details of theories are testable and which are not, given current observational constraints.

Within this framework of what has been observed and what can be observed in the future, there are two steps to studying exoplanets. The first one involves measuring the mass and radius to determine their bulk composition. 
The second step entails measuring whether the planet has an atmosphere and, if possible, determine its composition to connect this information to the bulk composition of the planet. Both steps ultimately inform us of the conditions of formation, which for a single planet or system may be limited, but with enough data, the hope is that these connections become robust and illuminating. 

The remaining structure of this review consists of revising current theories of formation (section \ref{formation}), that can be linked to the bulk composition (section \ref{rockyplanets}) and atmospheric properties (section \ref{volatileplanets}) of low-mass exoplanets. 
We will discuss separately an extreme class of planets, called the ultra-short period (USP) planets, with $\lesssim 1$ day orbital periods and scorching hot surfaces. Lastly, we will conclude by reviewing the lessons learned so far when looking at the very end stages of some stars that have already consumed all their fuel and have become white dwarfs. There is evidence that around some of these stars, solid bodies, including asteroid-like and Kuiper-like objects and small planets, are scattered into the dying star's tidal radius (maybe because of gravitational interactions with a more massive planet or with a passing star) at which point the strong gravitational force creates tides that disrupt the object and the resulting fragments are gradually swallowed by the white dwarf. 

\section{Composition Reflects Formation}
\label{formation}

The composition of a planet is the end result of a complex series of formation process that start with a proto-star surrounded by gas dust, and finish with a few remaining planets that underwent a dramatic growth history. The task is to parse out the formation pathways that lead to the fully formed planets. 

Before the advent of exoplanets, planet formation theories only needed to explain the solar system, characterized by the dichotomy of rocky planets in the inner solar system and gaseous and icy giants in the outer portion, with two debris belts at medium and large semi-major axis. Thanks to exoplanets and circumstellar disk observations, we now know those initial theories were highly biased. However, many aspects of solar system formation remain gold standards owing to the wealth of data available, including mass, radius, bulk chemical and isotopic compositions from studying the different solar system planets, meteorites, comets, asteroids and Kuiper belt objects, compared to that of exo-systems, for which we will never have such detailed cosmochemical information. Thus, our summary will focus on the lessons learned from the solar system formation, and we will describe connections made to extrasolar systems in both theory and observations. It will become apparent that generalizing from the solar system is not straight forward, given the numerous complex processes that happen during formation and the vast parameter space that has been poorly sampled with only 30 years of exoplanet discoveries. 

\subsection{Initial Phase: The Protoplanetary Disk Phase}
\label{sub:initial}
\subsubsection{Initial conditions of planet formation}
Stars are born in molecular clouds, that are concentrations of gas (mainly hydrogen) and interstellar dust. When a high-density region of the molecular cloud contracts under the effect of its own gravity, it forms a rotating protostar at the center surrounded by an envelope of gas and dust. Because of conservation of angular momentum, as the envelope contracts, a disk forms around the protostar. We call this a protoplanetary disk and in the case of the solar system it is referred to as the solar nebula. The protoplanetary disk is thought to contain about 1\% of the mass of the star and it is in this protoplanetary disk that planets are formed. 
This initial phase sets the stage for all subsequent processes, yet different aspects are still unknown. In particular, despite the tremendous progress, accomplished with the Atacama Large Aperture Millimeter/Submillimeter Telescope (ALMA) observations, the total amount of gas and solids in the protoplanetary disk, the size distribution of the solids, the disk radial and vertical structure and the evolution of the disk itself are highly uncertain while being key to understanding formation. For example, sources of uncertainties for the mass in solids comes from uncertainties in the dust microphysical (optical) properties and the optical depth. For the mass in gas, it is not possible to probe directly the emission of the main gas species H$_2$ because of faintness, so instead we rely on mass tracers like HD, different isotopologues of CO, and other molecular and atomic species, with the caveat that the interpretation of these observations depends on physicochemical models \citep{miotello2023}. Future protoplanetary disk surveys with ALMA and future facilities (like the Next Generation Very Large Array and the Atacama Large Aperture Submillimeter Telescope), probing fainter disks around a varied range of stars, combined with enhanced physicochemical models, should refine our knowledge of planet formation initial conditions
 \citep{miotello2023}.     

\subsubsection{Growth of solids}
Within this protoplanetary disk, the formation of planets begins.
The first step includes the condensation of submicron and micron-sized particles from the gas-rich protoplanetary disk, that subsequently grow to larger sizes: pebbles (mm-m sized), planetesimals (km to 100 km sized), embryos ($\sim$ 1000 km sized objects) and later planets. The composition of these first microscopic solids is determined from standard condensation sequence models that predict which compounds condense as a function of temperature, and partial pressure under thermodynamic equilibrium. Different lines of evidence support this view, including the depletion to varying degrees in the abundances of moderately volatile elements among the planets and signatures in chondrules in line with equilibrium conditions \citep{CondensationSS}. There is also evidence of the existence of presolar (or prestellar) submicron and micron-sized grains that originated from condensation within the molecular cloud phase and incorporated early in the protoplanetary disk \citep{testi2014}.

In broad terms, at high temperatures, Mg, Si, Fe, Al, Ti and Ca bearing minerals condense out ($\sim$600-1700 K), carbonates and carbonaceous compounds condense at lower temperature ($\sim$200-500 K), while ices (H${_2}$O, NH${_3}$, CH${_4}$) condense below $\sim$200K \citep{white2020geochemistry}. Thus, as the protoplanetary disk evolves, dust particles with different compositions condense out, the refractories being condensed in the inner portion of the disk and the ices condensing in the outer portions of the disk, beyond their respective condensation/ice lines. Ices do not condense in the inner portion of the disk because stellar irradiation limits the possibility of disk cooling \citep{Chiang:1999}. These submicron and micron-sized dust particles inherit the ``primordial signature" of the disk. Although details on how grains grow and may get sorted, as well as how the composition of the gas may change with the formation of planets, which we describe below, complicate the picture. 

At small sizes, grains can continue to grow and stick together thanks to microphysical processes (electric and Van der Waals forces). However, the intermediate growth phase, by which mm- or cm- size pebbles grow into km-sized planetesimals, is still one of the most outstanding questions of planet formation. 
Overcoming this barrier is a challenge because: 1) at sizes of $\sim$1 mm/cm, particles are not large enough to attract each other by gravity, but are too large for microphysical processes to enable growth, 2) their increased relative velocities produce more energetic collisions that lead to fragmentation, inefficient sticking or bouncing, rather than growth \citep{zsom2010}, and 3) they are no longer moving with the gas, but instead experience a gas drag that makes them spiral into the central star \citep{weiden1977} removing them before further growth can happen. But somehow planetesimals form.

Several mechanisms have been proposed to grow the grains beyond the mm/cm barrier, including low-velocity collisions from the tail of velocity distribution \citep{windmark2012}, highly-porous aggregates that increase the drift timescale and favor sticking when colliding \citep{okuzumi2009} and photophoresis \citep{Krauss2005}, among others.  Other proposed mechanisms circumvent the barrier entirely by invoking collective effects by which mm/cm-size  grains coalesce in mass to form planetesimals directly. These latter mechanisms include sedimentation and gravitational instability \citep{Goldreich1973}, turbulent concentration \citep{Cuzzi2008}, streaming instability \citep{youdin2002,youdin2005,johansen2007}, and the trapping of dust grains in pressure maxima \citep{johansen2004}. 

It is unknown if a single or a combination of these mechanisms are at play during formation, and different mechanisms might be dominating in different regions of the protoplanetary disk. It is possible, though, that the pathway to forming boulders from grains may lead to a chemical signature in the final planets providing a window for testing the different mechanisms via studies of exoplanets. For example, photophoresis \citep{Wurm:2013} and the streaming instability coupled to nucleation theory predict chemical sorting \citep{Johansen:2022}. These will be discussed in section \ref{rockyplanets}.

\subsubsection{From Planetesimals to Rocky Protoplanets}
\label{sub:pebbles}

Planetesimals can continue to grow within the gas disk via : 1) mutual collisions with other planetesimals or 2) pebble accretion, that continually supplies pebbles (mm to cm-size) to the accreting planetesimal. While planetesimals are not coupled to the gas disk owing to their sizes, pebbles are, and thus the aerodynamic drag from the gas reduces the velocity of the pebbles relative to the larger planetesimals, enhancing the cross section over which the planetesimals can accrete solids. This process rapidly accelerates growth \citep{Morby:2012, Chambers:2014} especially when pebbles are settled in the mid-plane and accretion proceeds in a 2D manner \citep{LiuOrmel:2018, OrmelLiu:2018}, until a point where the planet reaches its pebble isolation mass \citep{Bitsch:2018}. 

There is strong support for this process being responsible for the growth of the cores of giant and icy planets in the Solar System. This is because it solves the problem of formation timescale. Planetesimal growth from mutual collision at large distances is slow, and thus problematic for growing a core fast enough to accrete gas during the lifetime of the disk \citep{Dodson-Robinson:2009, Rafivok:2011}. Instead, pebble accretion onto 1000-km planetesimals is very fast, within the lifetime of the disk (1-10 My, \citet{Hillenbrand2008}), even at large distances \citep{Ormel:2010, Lambrechts:2012, Lambrechts:2014} providing a simple way to explain the giant planets in the solar system. 

As an extension, pebble accretion has also been studied as a process to grow the terrestrial planets to explain their masses and locations \citep{Lambrechts:2014, Levison:2015, Johansen:2021} and recently with greater chemical complexity with the goal of explaining isotopic constraints too \citep{Johansen:2023a, Johansen:2023b, Johansen:2023c}. Some of the challenges this mechanism presents are the quick timing of Earth's formation, within the lifetime of the solar nebula (4-5 My, \citet{Weiss:2021}) while Earth's core formation is known to take longer ($\sim$30 My \citet{Jacobsen:2005, Kleine:2017}), and an excess contribution of material from the outer portion of the disk \citep{Mah:2022} that violates the isotopic evidence that the Earth had about a 5\% mass contribution of solids from the outer disk \citep{Burkhardt:2021}. In addition, when considering the effect of fragmentation of pebbles, the inner disk produces silicate solids that are smaller ($\sim$ mm-sized) and more coupled to the disk which makes them inefficient to accretion onto a protoplanet. Instead, the icy pebbles outside the snow line have larger fragmentation sizes ($\sim$ cm-sized), and thus are less coupled to the gas enhancing their accretion efficiency onto the 
protoplanets in those regions \citep{Batygin:2022}.

The alternative way to grow the terrestrial planets is via collisions of planetesimals. While this theory had been around for long, it suffered from producing too massive a Mars when starting from an extended disk after gas dispersal \citep{Chambers:2001, Raymond:2009}. More recent work, however, seems to have solve this problem by invoking a narrow ring of planetesimals \citep{Hansen:2009, Nesvorny:2021}. Numerical simulations with a narrow ring reliably form Mars, Venus and the Earth in their location. Thus, perhaps the problem is growing the initial distribution of planetesimals in a narrow ring that can then grow into larger bodies via collisions \citep{Woo:2023}. Intriguingly, observations of disk structures with ALMA have consistently shown that ring-like structures are ubiquitous (see Fig \ref{fig:dsharp}).

\subsubsection{Effect of Disks' Structure and Composition}

\begin{figure}
    \begin{center}
     \includegraphics[scale=0.6]{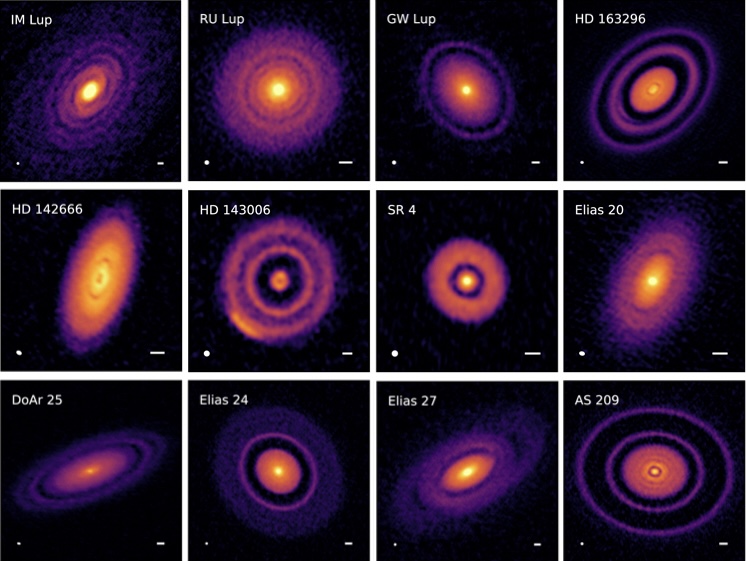}
     \caption{Dust emission in gas-rich protoplanetary disks observed by ALMA as part of the Disk Substructures at High Angular Resolution Project (DSHARP) \citep{Andrews2018}, showing a wide range of disk spatial features (rings, spirals, clumps). The scale bar at the lower right corresponds to 10 AU. Note that gas-poor debris disks, formed by the dust produced in planetesimal collisions, similarly show a wide range of disk spatial features (rings, spirals, clumps, large inner cavities, brightness assymmetries), spectacularly showcased in the new ALMA survey to Resolve exoKuiper belt Substructures (ARKS, Marino et al. in prep). Illustration credits: \citet{Andrews2018}.}
\label{fig:dsharp}
    \end{center}
\end{figure}

ALMA observations of large protoplanetary disks \citep{Andrews2018} have indeed shown that the dust spatial distribution features a rich structure, consisting of concentric, narrow bright rings where the dust is trapped and dark gaps where the dust has been depleted (see Fig. \ref{fig:dsharp}). Other features include large-scale spirals and azimuthal asymmetries. These substructures are compact and are found at a wide range of radii, from 5 AU to 150 AU. The ubiquitous narrow bright rings are thought to be due to dust trapped in axisymmetric gas pressure bumps \citep{Dullemond2018}. At these radial locations, planetesimals can form via sedimentation and gravitational instability \citep{Goldreich1973}. It has also been proposed that the rings and gaps could be due to dynamical interactions with already formed planets \citep{Zhang2018}. 

Disks extend beyond several icelines (rock condensation lines, water ice line, CO and CO2 icelines) that change the composition of the gas and the solids that condense out, and from which planetesimals and planets form. That is, beyond a condensation line, solids in the form of dust and pebbles trap the condensable compound, say CO or CO$_2$, depleting it from the gas, yielding a low C/H ratio. A radial drift of pebbles due to aerodynamic drag inside the condensation line will sublimate and release C, enriching the gas and enhancing the C/H ratio. Indeed, superstellar C/H ratios have been observed in several extrasolar giant planets \citep{Madhusudhan2011, Lavie2017} and in Jupiter and Saturn \citep{Owen1999, Atreya2005}. But this enhanced C/H ratio could also result from the contamination produced by planetesimal delivery or core-envelope mixing. From a single elemental ratio it is not possible to constrain the planet formation history but the combination of several elemental ratios can shed more light. 

\begin{figure}[htp]
\begin{center}
\subfloat{%
     \includegraphics[scale=0.11]
     {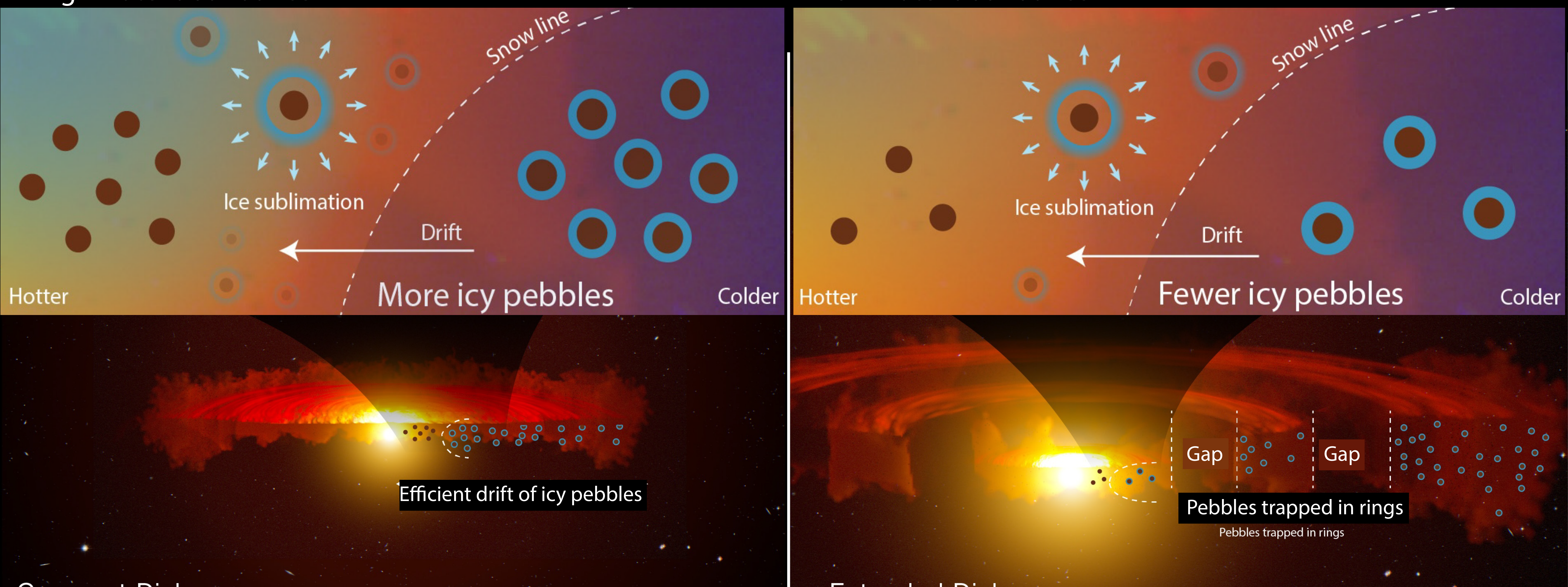}%
     }
     
\subfloat{%
     \includegraphics[scale=0.118]
     {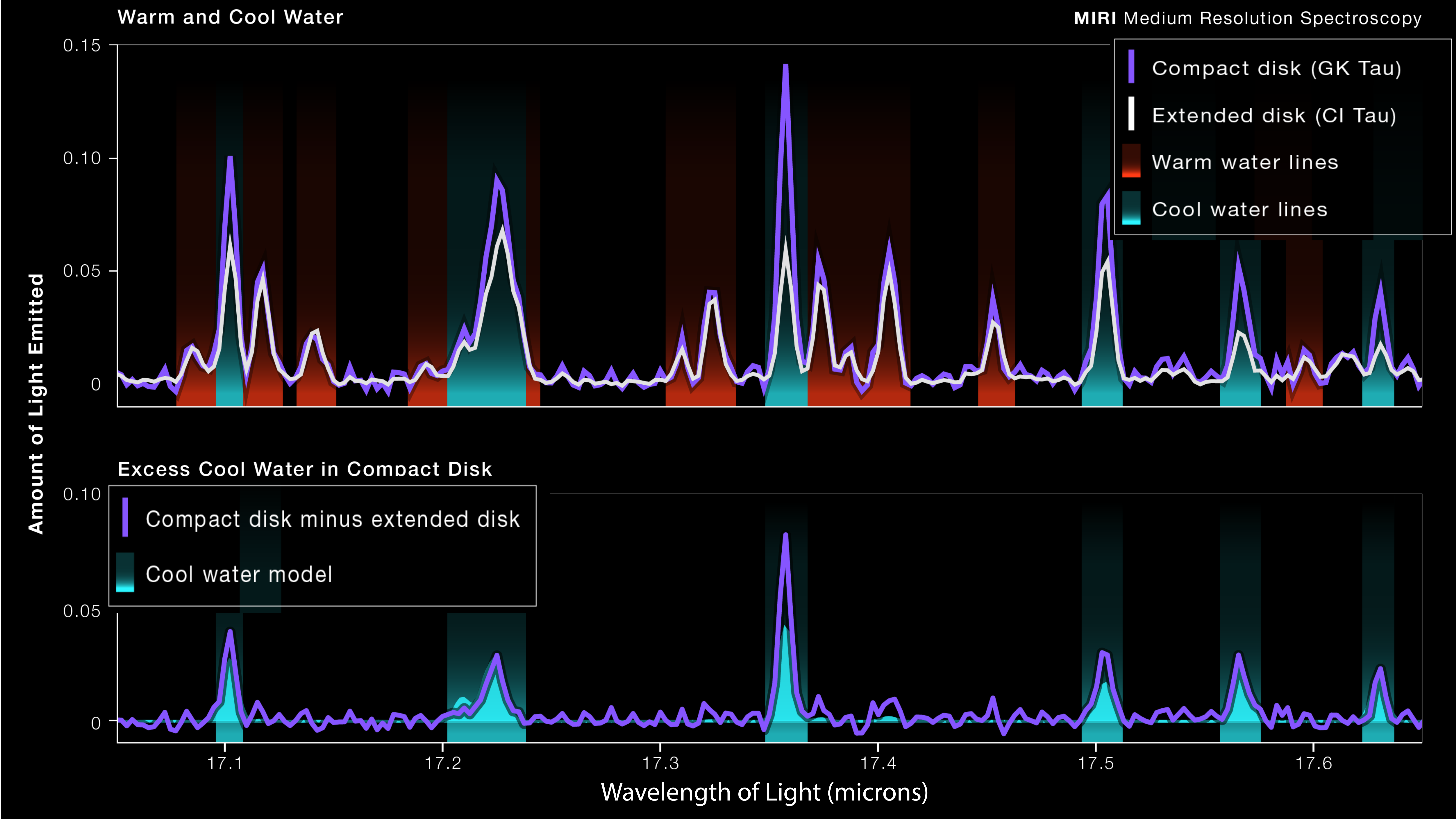}%
     }
     \caption{{\bf Top figure}: Schematic representation of the difference in gaseous water content in the case of a compact protoplanetary disk without gaps (left) and a more extended protoplanetary disk with gaps (right). When disk gaps are present, the drifting in of icy pebbles from regions beyond the iceline is impeded because pebbles get trapped in the rings; this results in a decreased water vapor enrichment in the inner disk region (1--10 AU) where rocky planets might be forming, compared to the case without gaps where a higher flux of icy pebbles cross the iceline and get their volatiles vaporized \citep{Banzatti2023}.  
     {\bf Bottom figure}: Spectra from the JWST Mid-Infrared Instrument showing water vapor emission in a a compact protoplanetary disk without gaps (purple line) and in an extended disk with gaps (white line), showing excess cool water in the former, as expected if the water enrichment is the result of pebble drift. Top figure illustration credits: NASA, ESA, CSA, Joseph Olmsted (STScI). Bottom figure illustration credits: NASA, ESA, CSA, Leah Hustak (STScI). Science credits: Andrea Banzatti (Texas State University). 
}
\label{fig:PebDrift}
\end{center}
\end{figure}

Any mechanism that can stop the drift of solids, such as the formation of pressure bumps or the presence of massive planets that can open a gap in the disk, would erase this chemical gas signature near icelines. In addition, this would impede the delivery of volatiles to the inner disk affecting the volatile budget of planets forming at those locations. Figure \ref{fig:PebDrift} shows recent JWST observations of protoplanetary disks with and without gaps (two disks of each type were observed). It was found that the two disks without gaps have an excess emission in the low-energy levels of water (corresponding to T $\sim$ 170-400 K and an equivalent emitting radius of 1--10 AU), compared to the two disks with gaps, in agreement with this impeded/unimpeded icy pebble drift scenario \citep{Morbi2016,Banzatti2023}.

\subsubsection{Effects of Giant Planet Formation}
The observed median protoplanetary disk lifetime is 2.5 Myr, but some disks can last up to 10 Myrs \citep{Hillenbrand2008}. Classical giant planet formation theory posits that once the rocky core has grown comparable in mass to that of the gas it gravitationally holds, there is an exponential accretion of gas leading to giant atmospheres \citep{Pollack1996}. If the core does not grow quickly enough, a bare compact planet is left. The intermediate case arises when the mass of the envelope is less than that of the core, preventing runaway gas accretion (like Neptune and Uranus). The almost completely formed gas planet may continue to accrete solids onto its atmosphere in the form of pebbles or planetesimals, affecting the metallicity of the atmosphere if settling onto the core is incomplete. 

The formation of gas giant planets can have a strong influence on the dynamics and stability of a disk and thus, on the accretion, growth and survival of other bodies in the system. Whether they migrate during the gas phase \citep{Walsh2011} or cross dynamical resonances later \citep{gomes2005}, the effect is to cause high eccentricities and orbital inclinations\footnote{The tilt of an planet's orbit around the star.} in the rest of the bodies that leads to increase collisions, changes in orbital parameters, and even ejections of small and even larger planets \citep{Nesvorny2012}\footnote{Under the Nice planet formation model \citep{NiceModel} that captures many features of the formation of the Solar System, there is a 20\% chance that another Neptune-like planet was ejected during formation \citep{Deienno:2017}.}. By parsing out the differences and similarities in composition and orbital properties of exoplanetary systems with and without gas giant planets, we stand to better understand planetary system formation and evolution scenarios. 

\subsection{Giant Impact Collision Phase}
\label{sub:collisions}

After the bulk of the gas is gone, giant planet formation comes to an end but the collisional growth of other planets may continue, if they are sufficiently close to feel their mutual gravitational interactions \citep{morbi2012}. 
At this point, a process that might have started in an orderly fashion quickly transitions into a fairly chaotic state. If giant planets have formed, they will clear their feeding zones of planetesimals, accreting some and ejecting others out of the system. This balance between ejection or accretion depends on the ratio between orbital velocity and escape velocity from the surface of the scattering planets. In the inner system accretion is the most likely fate. Smaller-mass planets may not clear their feeding zone and continue to grow on longer timescales via collisions (10-100's of My). During oligarchic growth\footnote{Oligarchic growth describes the runaway accretion of the largest planetesimals into embryos of similar masses through the acquisition of smaller planetesimals leading to the formation of two populations, that of the oligarchs and that of the smaller planetesimals. The mass gap between these two is a few orders of magnitude \citep{Raymond:2022}.} a number of embryos grow, their radius of influence (e.g Hill radius) expands, causing mergers and giant impacts. When the total mass in the newly formed planets is comparable to the mass in planetesimals, the dynamical friction by the planetesimal population is no longer able to damp out the dynamical excitation triggered by the growing planetesimals and embryo planets, marking the beginning of the high velocity giant collisions phase \citep{Goldreich:2004, Kenyon:2006}.

These giant collisions, which involve bodies of similar sizes impacting at high speeds, are highly energetic, global-scale events that have the potential to alter the composition of planets, both their atmospheres, and even their bulk chemistry. Within the same epoch, the type of process dominating in the inner disk may be different than that of the outer disk, given the longer formation timescales at longer semi-major axes. While planetesimals may be cleared in the inner part of the disk, in the outer part they may still be colliding. The swarms of planetesimals will interact with the growing planets and this can cause their migration and trigger episodes of dynamical instability and orbit readjustment \citep{gomes2005}. During these gravitational instabilities, planetesimals can be scattered out or scattered in. In the latter case, they can become a source of volatiles to the inner region \citep{morbi2000,raymond2009} where the collisional growth of planets is still continuing. 

Irrefutable evidence of giant impacts comes from the formation of the Moon after a mars size body (Theia) collided with Earth. Compelling evidence for at least one more giant impact on Earth comes from noble isotopic measurements \citep{Tucker14}. The values for $^3$He/$^{22}$Ne in the upper mantle exceed that of the Sun, implying there must have been a mechanism to remove Ne. Additionally, the low ratios of $^3$He/$^{22}$Ne in samples from deep mantle plumes suggest an ingassing of nebular He and Ne that was somewhat preserved throughout Earth's evolution. \citet{Tucker14} suggested that the increase in $^3$He/$^{22}$Ne isotopic ratio is due to the preferential outgassing of Ne from an enriched magma ocean owing to the factor of 2 larger solubility of Ne compared to He. The low $^3$He/$^{22}$Ne ratios of deep mantle plumes suggest the existence of partial magma oceans after at least two giant impact collisions. 

Other observational evidence of giant collisions come from debris disks. These are dusty disks produced by collisional processes that take place as the planetary systems form and evolve and include pseudo-steady state collisional activity, resulting in the production of dust during long periods of time, punctuated by stochastic events associated with large individual collisions, resulting in short-lived dust production \citep{Wyatt2008, Beichman2005, Moro2013, Meng2014, Su2019}. 

Of particular interest in the context of giant collisions are the ''extreme'' debris disks that were first identified by the space infrared telescope {\it Spitzer}. They have been interpreted as the result of collisional activity during terrestrial planet formation, or during the dynamical instabilities that followed after the gas dispersed. The location of these events in the innermost region of the planetary system is inferred from the high temperature of the dust observed during these brightening events and from the short timescale of the variability. Indeed, some ''extreme'' debris disks \citep{Meng2014, Su2019} show significant variability (as much as a factor of two over a period of days to months), consistent with violent impacts involving large asteroid-size bodies; some also show strong mineralogical features in their mid-infrared spectra indicative of fine silica grains thought to originate from the condensation of the silica gas that is produced in violent collisions as the rocky material vaporizes (e.g. the Moon-forming event). Because this silicate gas has a short lifetime $<$ 100,000 years or 100-1000 years in a circumplanetary disk \citep{Lisse2009}, its detection could help identify planetary systems experiencing recent violent collisions. 
Long-term monitoring of these "extreme" debris disks with the space infrared telescopes {\it JWST} and {\it Roman}, together with archival data from {\it Spitzer} \citep{Su2022}, focusing on the composition of the debris dust and gas, the rate of large collisions, and the timescale for debris clearing, will allow us to better understand the chemical make up of solid bodies in the terrestrial planet region of planetary systems and the role of giant collisions in planet formation. 

\subsection{Volatile accretion}
\label{sub:volatileaccretion}

Understanding how Earth and the Terrestrial planets acquired their volatile contents, especially water and carbon, is crucial to knowing how Earth became habitable. Water is important, for obvious reasons, as all life forms on Earth require it for their survival. However, Earth's bulk water content is not particularly high, in fact it is intermediate between water-poor and water-rich asteroids, and it is much lower to that inferred in some exoplanets (ie. LHS~1140~b seems to have ~10-20\% water, \citet{Cadieux:2024}). Carbon is also important on Earth to provide, in conjunction with water, a feedback mechanism on climate to keep it balmy for billions of years, the so-called C-Si cycle \citep{Kasting2003}. However, too much carbon can be detrimental for a planet as carbon outgassing can easily outpace carbon sequestration leading to a broken imbalance \citep{Valencia2018}. Thus, understanding how planets acquire volatiles is important and exoplanets may offer clues. 

There are two ways in which volatiles have been proposed to be accreted: absorbed from interaction of a magma ocean in contact with an atmosphere, or delivered by volatile rich bodies via collisions.  

\subsubsection{Absorption from Nebular Gas via a Primordial Atmosphere}
\label{subsub:absorptionnebula}

As seen before, planetesimals and even planetary embryos grow in the presence of the solar nebula in the protoplanetary disk. If they remain solid, the interaction with a gaseous medium would be negligible, but if there is any melting, there could be ingassing into the growing protoplanet. Pioneering work by \citet{Hayashi1979,Mizuno1980,Nakazawa1985} found that protoplanets within a nebular gas would gravitationally bind atmospheres that thermally blanket any radiation from the planet melting its surface. Despite gaining little traction for decades, this idea has more recently been revisited to understand the origin of water and volatiles on Earth through the interaction of H in the nebula with oxides from the magma ocean \citep{Ikoma2006, Olson2019, Gaillard:2022}. More recently, \cite{Young2023} considered nebular ingassing with a detailed chemical model and proposed not only that it is a source of water for Earth but a mechanism to reduce the density of the core owing to the incorporation of H, as well as O and Si at relevant conditions. Partitioning coefficients at the relevant pressures and temperatures during core formation would determine the D/H isotopic signature in this scenario, and serve as a test for this model. 

Ingassing onto protoplanets could happen for masses as low as $\sim0.3 M_E$ \citep{Ikoma2006}, meaning that possibly the precursors of Earth and Venus could have experienced ingassing, whereas Mars and Mercury would have been dry. Strong support for the role of ingassing on Earth is evidenced by noble isotopes. Deep mantle plume sources show $^{20}$Ne/$^{22}$Ne ratios similar to the Sun and to those expected in the protoplanetary disk, and distinct from CI chondrites or solar wind implantation \citep{Williams2019}. Also, the low concentration of $^3$He/$^{22}$Ne in deep mantle plumes is consistent with ingassing from nebular gas \citep{Tucker14}. 

The study of magma ocean interactions with a primordial atmosphere has been naturally extended to rocky exoplanets \citep{Kite2021,Sossi:2023,Bower:2022, Charnoz:2023, Peng2024}, especially in light of the fact that there are many super-Earths that are so highly irradiated that they should have a permanent magma ocean (see section \ref{USP}). Notably, different JWST campaigns are targeting these scorching planets opening a window to understanding atmosphere-magma ocean chemical and thermal exchange. 

\subsubsection{Delivery by Collisions}
\label{sub:deliverycollisions}

Based on isotopic signatures (in O, Cr, Ti, W and Ni), meteorites are generally divided into two groups: carbonaceous, that are mostly water abundant, and non-carbonaceous that are mostly water poor. This isotopic dichotomy supports the idea that two distinct unconnected chemical reservoirs in the solar system were established early on (at $\sim$1 My, \citet{Kruijer2017}). 
This picture is consistent with early's Jupiter's formation (within the first Myr after the formation of the CAIs) when the protoplanetary disk was still hot \citep{Morbi2016, Kruijer2017}. Alternatively, it could also be explained by different timing of planetesimal formation versus location of water ice line \citep{Lichtenberg2021}. In the former scenario, it is thought that Jupiter formed between 3-5 AU, close to the snowline, and once it had grown to 20-30 Earth masses, it separated planetesimals into the two reservoirs, the volatile-rich, beyond its orbit, and the volatile-poor, inside its orbit. Jupiter interrupted the drift of the volatile-rich planetesimals into the terrestrial planet region \citep{Morbi2016, Kruijer2017}. Also, from the trend in moderately volatile elements, the Terrestrial Planets appear to have formed mostly dry \citep{CondensationSS}. 

Within this context, a popular idea is that water on Earth came from the wet outer regions of the disk via collisions. Different dynamical scenarios that can explain other solar system features are consistent with this idea. For example, one scenario called the Grant Tack model \citep{Walsh2011} invokes the migration of Jupiter within the disk inwards until the growing Saturn catches up to it into a 2:1 resonance at which point they reverse directions forcing Uranus and Neptune to migrate outward as well. Jupiter sweeps the asteroid belt as it moves inwards and outwards. This process explains the small size of Mars by restricting its feeding zone, and the chemical and dynamical signatures of the asteroid belt \citep{Jewitt2015}, while also leading to the scattering of icy planetesimals formed beyond the snowline (in the 5--13 AU region) into the inner solar system after most of the terrestrial planet growth would have already finished (10-100 Myr). 
The majority of these wet implanted planetesimals would be the present day ice-rich asteroids, but some of them could have delivered water to the terrestrial region. Assuming they were 10\% by mass water rich, the wet planetesimals would deliver 1-2\% of the Earth's mass in water \citep{OBrien2014}, which would easily account for Earth's bulk water budget. This volatile delivery scenario is in agreement with the similarities found in the isotopic ratios of $^{14}$N/$^{15}$N and D/H between the Earth, Mars, and the present-day ice-rich asteroids. In contrast, these ratios are different in comets, with the caveat that they are not well know for a wide sample of comets and other objects in the outer solar system (see review by \citet{Alexander2018} and references therein). 

The early timing of this volatile accretion also explains the presence of water in the Moon in minerals, volcanic glass and basalt inclusions, with D/H ratios similar to Earth. Basically, the water was inherited from the Earth as the silicate vapor cloud that was released after the Moon-forming giant impact, and from which the Moon condensed out. The timing of events also implies that part of Earth’s water survived the moon-forming impact. The amount of volatiles that get degassed \citep{Abe1985,Abe1988,Matsui1986a,Matsui1986b} and ingassed \citep{Elkins2008} after collisions are the results of a chemical exchange between the magma ocean that is formed or was already present, and the atmosphere, that either existed or is evaporated upon impact.  

In exoplanets however, there could be different dynamical circumstances that could lead to volatile delivery via collisions at different epochs. For example, on Earth the relative high abundance of siderophile elements in the mantle are evidence of delivery of material by collisions after the core formed. On Earth this delivery is very small, accounting for only 0.5--1.5\% of the Earth's mass \citep{Dauphas2002} including a negligible amount of water. However for other systems, depending on the dynamical circumstances that led to these instabilities, these impactors could deliver higher abundances of volatiles.

Impact delivery, while having been historically more adopted, is not the only way to explain Earth's water budget. Recent work has proposed that the H content of enstatite chondrites, which are non-carbonaceous meteorites and expected to form in the inner part of the disk due to their reducing conditions, can account for the entire water budget of Earth \citep{Piani2020}. This scenario is especially compelling given the fact that Earth resembles enstatite chondrites in at least 15 isotopic systems \citep{Javoy1995,Javoy2010,Sikdar:2020}, and reinforces the emerging idea that Earth's precursors might have been mostly enstatites \citep{Dauphas2017}.

\paragraph*{Volatiles in the Inner Region of Extrasolar planetary systems}

There is tentative evidence from debris disks observations that volatiles can be accreted via collisions with planetesimals originating in the outer planetary system, well after the gas in the protoplanetary disk has dissipated. The tentative evidence is based on: 1) The presence of cold gas in the outer region of debris disks observed by ALMA \citep{Lieman-Sifry2016, Moor2017, Marino2017, Smirnov-Pinchukov2022}, pointing to the existence of reservoirs of volatile-rich planetesimals that are ougassing via collisions or crossing the ice line \citep{Hughes2018, Kral2017, Cataldi2023}. 2) The presence of planets embedded in the disks observed by either direct imaging \citep{Meshkat2015} or inferred from the presence of gaps in the dust disk structure, thought to be created by planets \citep{Pearce2022}; these planets would be able to trigger dynamical instabilities that could scatter planetesimals from the outer into the inner planetary system. 3) The presence of transient hot atomic gas ($\geq 1000$K) near the star that changes within a matter of hours or days \citep{Rebollido2018, Rebollido2020, Rebollido2022, Roberge2008, Roberge2006, Roberge2014} thought to arise from gaseous clumps passing in front of the star that form from the sublimation of cometary bodies scattered from the outer system. Such origin would be consistent with the presence of warm carbonaceous grains found in the inner regions of some CO-rich disks, like in the case of eta Crv \citep{Lisse2012}. This scenario, in which there is a delivery of volatiles from the outer to the inner region of a planetary system needs to be confirmed and this is the goal of ongoing studies with JWST that search for CO, water and other volatiles in debris disks in the inner, outer and intermediate disk regions. 

\subsection{Volatile Loss}
\label{volatileloss}

In addition to volatile accretion or delivery, volatile loss determines the final budget of a planet. 
There are different mechanisms for volatile loss. Here we will describe the two most important ones in the context of extrasolar planets: hydrodynamic evaporation and loss from collisions. 

\subsubsection{Hydrodynamic Evaporation}
\label{subsub:hydrodynamic}

Large scale mass loss can be driven by hydrodynamic escape when heat deposited in the upper atmosphere drives outflow and two mechanisms can facilitate this: photoevaporation and/or core-power mass loss. Photoevaporation entails heating from the absorption of extreme UV rays from the parent star that interact predominantly with H driving hydrodynamic mass loss \citep{Yelle2004,GarciaMunoz07,Erkaev07, MurrayClay09, OwenWu13}; these hydrogen winds create a drag that can lift heavier compounds removing them as well \citep{Tian15}; the active phase of the star, where its bolometric flux is at its highest and can easily drive photoevaporation, varies between 100 My for Sun-like stars to 1-2 Gy for small M dwarfs \citep{Selsis2007}. The second mechanism, core-power mass loss, involves the heating of the upper atmosphere with infrared radiation from the cooling of the solid planet from below and the bolometric flux from the star \citep{Ginzburg18, Gupta_Schlichting19}. While both processes are expected to happen at the same time, which one dominates depends on whether the sonic point from which hydrodynamic escape is launched is deeper than the penetration layer of the extreme UV rays, or not \citep{Bean2021}. In both cases, the closer to its parent star and the lighter the planet is, the larger the atmospheric loss, such that hydrodynamic flow can efficiently strip highly irradiated volatile planets of their atmospheres leaving them bare. 

In fact, one of the most important features in the population of low-mass exoplanets is the existence of a ``radius valley". \citet{Fulton17} found there are two distinct populations centered around $1.3 R_E$ and $2.4 R_E$ for planets around FGK stars, with very few planets within the valley. The compact population is presumed to be of rocky planets, while the larger planets are volatile mini-Neptunes. This radius valley exhibits a positive correlation with stellar irradiation (instellation) congruent with sculpting via either photoevaporation \citep{OwenWu13, Lopez2018}, or core-powered mass loss \citep{Ginzburg18, Gupta_Schlichting19}. 
This radius valley persists in the population of planets around M dwarfs \citep{Cloutier20}, albeit perhaps more populated \citep{Bonfanti:2024}. Controversy arises on the correlation with stellar irradiation or period, ranging from a negative correlation \citep{Cloutier20} consistent with accretion under a gas-poor environment \citep{Lee_Chiang16,Lopez2018}, a positive correlation similar to Sun-like stars \citep{VanEylen:2021} or an almost flat correlation \citep{Bonfanti:2024} suggestive of a combination of processes. This ``radius valley" and the differences among different stars holds important clues as to how planets form and is being intensely explored by different observational campaigns from the ground and space. 

In contrast, in the solar system the evidence for hydrodynamic escape comes in the form of isotopic fractionation. For example, the D/H ratio of Venus \citep{Donahue1982} has been explained due to water dissociation and subsequent hydrodynamic escape \citep{Genda2005}. The Earth does not show evidence of isotope fractionation characteristic of hydrodynamic escape, which implies that, if this loss mechanism occurred, it was only important in the embryonic stage when the atmosphere was richer in H and that subsequent atmospheric accretion and loss mechanisms deleted its geochemical signature \citep{Schlichting2018}.  

\subsubsection{Atmospheric loss due to collisions}
\label{subsub:losscollisions}

Collisions do not always and only deliver material, they can also be an efficient source of mass removal from the impacted body. In a simple model of atmospheric loss (like that of \citet{Schlichting2018}), a collision between two planetary bodies leads to local atmospheric loss caused by the expansion of a plume at the impact site that, at minimum, affects the atmosphere vertically above the impact and, at maximum, affects all the atmosphere above the tangent plane at the impact site. \citet{Schlichting2018} discuss three regimes that predict the fate of the atmospheres: giant impacts that in addition to a local acceleration of the atmosphere, also create a strong shock front that propagates through the entire planet causing ground movements that, depending on their strength, can cause atmospheric loss globally \citep{Genda2005,Stewart2014,Schlichting2015}, and can be significantly more effective at atmosphere removal in the presence of an ocean \citep{Genda2005}; impacts from large planetesimals that can only eject the local atmosphere \citep{Schlichting2015}; and small impactors that eject a fraction of the atmospheric mass above the tangent plane. Small impactors are the most efficient at eroding the atmosphere \citep{Schlichting2018}, given that there is no energy being wasted in generating a shock through the planet. Because the size distribution of impactors generally follows a power-law, with orders of magnitude more smaller bodies, planets in general experience a small number of giant impacts that can generate magma oceans, and a larger number of smaller planetesimal collisions that can efficiently erode the atmosphere \citep{Schlichting2018}. 

\subsection{Concluding remarks on planet formation}
\label{formationconclusions}

The end result of planet formation are planets with a composition that has been affected by a wide range of processes. Timing of the protoplanetary disk dissipation plays a critical factor in the types of planets that form, their dynamical evolution and their collisional history. Therefore, the range of protoplanetary disk lifetimes will also result in a range of planet and planetary system characteristics. As described above for the case of the solar system, the current most favored scenario implies that the early formation and subsequent migration of Jupiter and the other giant planets altered the availability and composition of solids in the terrestrial planet region, strongly determining the size and composition of the terrestrial planets and their subsequent collisional history, differentiation, volcanism and tectonics. The situation might have been very different in other planetary systems where the early presence of a giant planet in the inner region may not have occurred. This could have allowed water worlds, super-Earths and/or mini-Neptunes to accrete.

The solar system is the result of a very particular formation history. A broader perspective on planet formation can be gained by studying extrasolar planets, featuring a very wide range of planet characteristics and planetary system architectures. This review focuses on how the study of the extrasolar planets' compositions can help us understand planet formation. This is not an easy feat given the scarcity of data and the complexity of planet formation pathways.

\section{Planets in the Present}
\label{planetspresent}

Having reviewed the major ingredients of planet formation we now turn to the lessons learned from exoplanets and future perspectives. The diversity seen in exoplanets in terms of their characteristics points to a diverse formation and dynamical history. 

The composition of a planet is studied with interior structure models that assume a two- or three-layered structure, with the volatile envelope above a solid silicate mantle, above an iron core. The more sophisticated models allow for mixture of these reservoirs when need be. 

\subsection{Rocky Planets}
\label{rockyplanets}

\begin{figure}
     \centering  
     \includegraphics[scale=0.85]{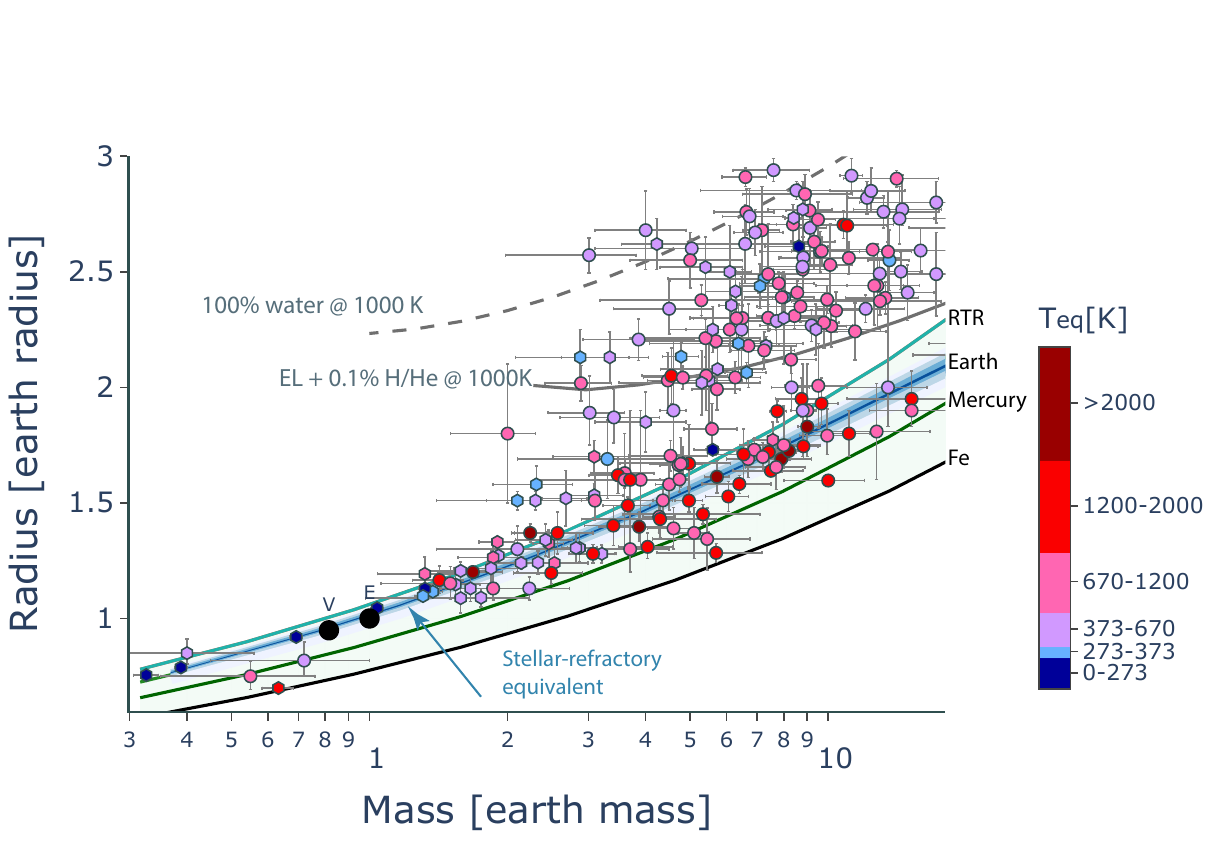}
     \caption{Exoplanets with masses and radii errors less than 20\%. Colors correspond to equilibrium temperatures calculated with bond albedos of 0.3 and redistribution factors of 4. Color scheme is the same as for Fig. 1. M-R relations for a pure iron planet, an iron-rich planet like Mercury (Fe=74\% by weight), a planet like Earth (Fe=32\% by weight), and planet voided of iron and made of MgSiO3+MgO that depicts the largest size a rocky planet can be (rocky threshold radius, RTR). Planets above the RTR necessarily have volatiles. The blue shaded region shows the M-R relations of the refractory ratios of stars. Only planets in this narrow region may be explained by a primordial origin. For reference solar system planets are shown as black symbols, an Earth-like composition below a 0.1\% H/He atmosphere at 1000K is shown in solid  line, and a 100\% steam planet at 1000K is shown in dashed line. Any planet above this latter line with the corresponding temperature, or cooler, requires H/He. }
     \label{fig:MR}
 \end{figure}

As of today, according to NASA Exoplanet Archive, there are 162 planets with R $\leq$ 2 R$_E$ and M $\leq$ 10 M$_E$, or only 129 planets where one of those values is not an upper limit only. From these two measurements alone, it is not possible to determine with full confidence that a planet is rocky owing to the degeneracy in composition where a planet can have a smaller mantle, and/or core and have a volatile content (either a gaseous envelope or water/icy layer). Of high interest is to confirm or rule out the presence of an atmosphere by observing their phase curve (see section \ref{ExoplanetObservations}). Already for four planets, LHS 3844b \citep{Kriedberg19}, Trappist-1 b \citep{Greene2023} and Trappist-1 c \citep{Zieba2023}, GJ~367~b \citep{Zhang2024}, the observations are consistent with a rocky nature. For all other planets that fall within the rocky region (see Fig. \ref{fig:MR}), there is always a chance that they have a volatile envelope, however, the more compact the planet is, based on formation arguments, the more likely it is made of rocks. It is important to clarify that Earth with its water budget is considered to be rocky, given that its volatile content does not change its structure in a measurable way from an exoplanet perspective. 

A mass-radius (M-R) measurement for a rocky planet constrains the amount of heavy to light material owing to the fact that these have different densities \citep{Plotnykov24}. Given the abundances in stars, the heavy materials are mostly iron (Fe) and nickel (Ni), and the light materials are silicon (Si), magnesium (Mg), and oxygen (O). Given the comparatively small uncertainty in Ni/Fe in stars, and assuming the planets follow the stellar ratio [which is a common assumption for Earth, \citet{McDonough_Sun95}], the combination of M and R measurements constraints the amount of iron in the planet. To be more precise, taking into account degeneracy with volatile content, a M-R measurement yields the minimum iron content of a rocky planet. The radius of a planet is much less sensitive to the degree of differentiation (i.e. where the iron is found, whether in the mantle or core), the amount of alloy in the core displacing iron, and the ratio of Mg/Si \citep{Plotnykov20}. This means that even for the best mass and radius errors in rocky exoplanets of $10\%$ and $2\%$ respectively, these details in interior structure have no effect on the inferred iron-mass fraction.

\begin{figure}
     \centering  
     \includegraphics[scale=0.70]{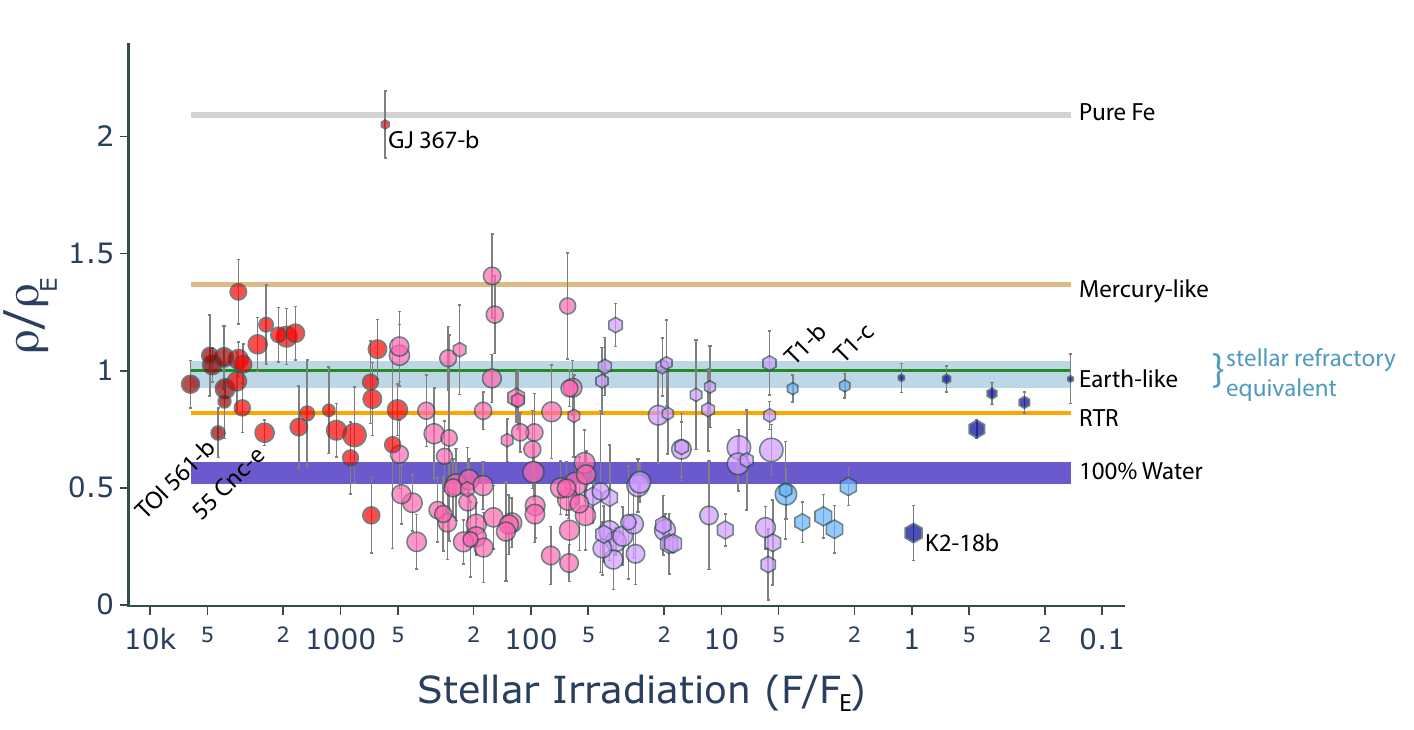}
     \caption{Normalized density of low mass exoplanets (with M and R errors below 20\%) compared to the density of the Earth, Mercury, a pure Fe composition and a pure silicate oxide one (same as Fig 6), and 100\% condensable water, as a function of normalized incident stellar irradiation with respect to that received by Earth from the sun. Colors follow the same scheme as Fig. 1. The light blue band shows the density of planets made out of the same refractory relative abundances as of stars within 2-$\sigma$ values. Only planets in this narrow region may be explained by a primordial origin. Planets with densities below that of condensed water require a gaseous envelope. Planets with densities below that of RTR necessarily have volatiles. The thickness of the lines shows the differences in density once the planetary mass and hence pressure effects are considered. The lower bound corresponds to small planets and the upper bound to massive planets. Planets labeled are those that have JWST measurements. }
     \label{fig:rho}
 \end{figure}

Exoplanet data (Fig. \ref{fig:MR}) shows that there is a spread within the composition of alleged rocky planets. Another way to compare composition is to use either the uncompressed density, or the density normalized to that of an Earth-like planet at the same planetary mass. The latter does not remove completely the effects of compression on density, but it is a practical comparison space. Figure \ref{fig:rho} shows the different class of planets according to their normalized bulk density as a function of irradiation including those that have been observed with JWST. The lack of planets at very high irradiation levels and low densities, despite the favorable bias to detection, is consistent with atmospheric evaporation theories. The shaded light blue region depicts the refractory abundances of the stellar population at the 2$\sigma$ level ($\rho_\star^{+2\sigma}=1.037 \rho_E$,\quad$\rho_{\star}^{-2\sigma}=0.934 \rho_E$).

\citet{Plotnykov20} inferred the core mass fraction (CMF) of planets\footnote{At typical M-R errors in exoplanets both CMF and iron-mass fractions are equivalent \citet{Plotnykov24}]} with good M-R data and showed that the CMF of rocky exoplanets has a wider range compared to that of stars in a population sense by a factor of $\times 3$. 
One-to-one comparisons \citep{Plotnykov20,Schulze21, Adibekyan21, Wang2022} have been suggestive but inconclusive as to whether the stars and planets have the same relative abundances of refractory elements (or refractory ratios) in support of a primordial origin. The reason is mostly because of large error bars in planetary mass, but in some cases in stellar composition. This primordial origin works for Earth, likely for Venus and Mars, but not for Mercury. Thus, it is of extreme importance to understand how common it is for planets to end up with a similar ratio of refractory elements as their star, or if there is any processing happening during formation that yields different results. 
In more detail, \citet{Adibekyan21} conducted planet-star direct comparisons which suggested a positive correlation between iron content of the host star and that of its planet, including a distinct population of super-Mercuries. However, the data used by them and all other works suffered from not ensuring consistency between the composition of the star and the stellar mass and radius inferred, which feeds into the results for planetary mass and radius. 

\begin{figure}
     \centering \includegraphics[scale=0.5]{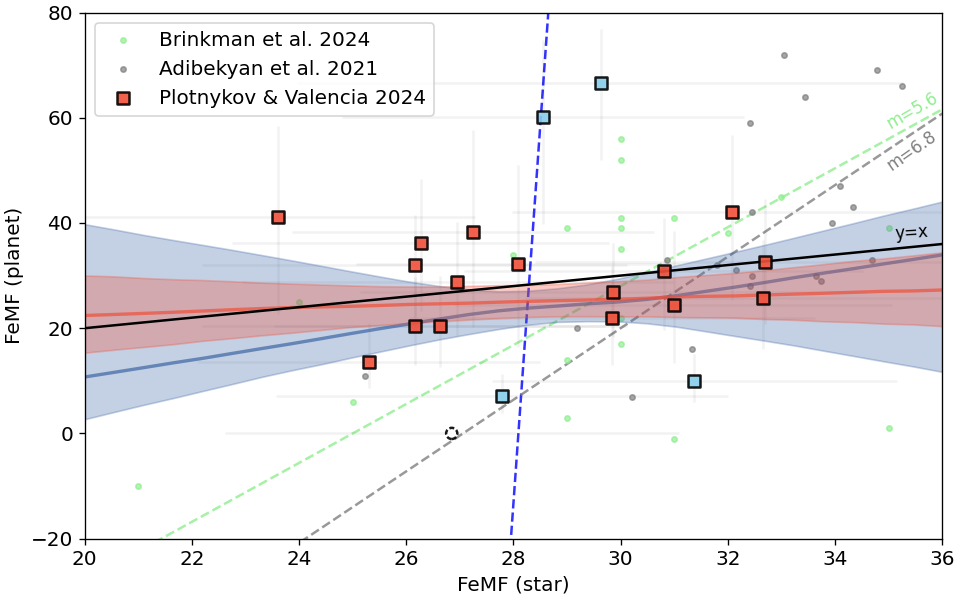}
     \caption{Iron content of rocky planets and their host stars as calculated by different group, where Fe-MF stands for iron-mass fraction. Dashed lines show the correlations for the data from \citet{Adibekyan21} (grey), Brinkman et al 2024 (green) and \citet{Plotnykov24} (navy blue) using all the sample size, inferred by using the orthogonal distance regression \citep{Feigelson1992}. Squares show the data from Plotnykov and Valencia 2024. The open circle symbol represents the underdense 55 cnc-e planet which recently has been shown to have an atmosphere \citep{Hu:2024}, making the core inference less reliable. Shaded regions shows the inferred correlations using a Bayesian framework considering all planets (light blue shaded region includes blue squares and red square planets) or those that have more reliable data (red shaded region includes only red square planets). There appears to be little to no correlation between the iron content of planets and their host stars.} 
     \label{fig:corr}
 \end{figure}

In response to this shortcoming, recently two different groups reanalyzed stellar data ensuring consistency (Ross et al. \emph{in prep}, and Polanski et al. \emph{in prep}) and provided updates to planetary masses and radius. These updated datasets in general agree within one sigma values but differ in significant ways for some planets compared to the data used by all previous works \citep{Plotnykov20,Schulze21, Adibekyan21, Wang2022}. Subsequently, two different groups (Brinkman et al, 
\emph{submitted}, Plotnykov and Valencia, \emph{in prep}), used these updated data sets to rederive any correlations.
With different statistical approaches, these updated data seem to indicate there is no correlation between the stellar and planetary iron contents. 
However, the combination of narrow stellar compositions, small data sets and large planetary errors make any inferences very limited (see Fig. \ref{fig:corr}). It also means that a few outlier planets can skew the results and should be carefully followed up.
Furthermore, while this data is consistent between planets and their star, it still suffers from being heterogeneous in the sense the transit and RV data were collected by different instruments and analyzed by different groups that could introduce biases \citep{Teske2021}.

\paragraph{Super-Mercuries}
As evident from the data, there are two very important groups within the population of compact planets: the iron-enriched rocky planets, or super-Mercuries, and the iron-poor rocky planets or super-Moons. How prevalent are they? Given that for a \emph{given radius}, radial velocity detections are biased to larger masses, observations are biased to detecting super-Mercuries with larger amplitudes, rather than super-Moons. 
In addition, some of these very compact planets may not be so compact. That is, if they are sufficiently close to their star, they are probably distorted due to tidal forces and their measured transit radius is smaller than their actual undisturbed radius. This is because by being tidally distorted, their elongated radius points towards the star, the intermediate radius points along the orbit, and the shorter radius is the polar radius. Given that the transit radius is sensitive to the area that blocks the light, namely the intermediate and polar radii ($R_{\rm transit}=\left(R_{\rm int} R_{\rm polar}\right)^{1/2}$, \citet{Fei:2024}), the undisturbed radius is larger than the transit radius. This seems to be the case for TOI-6255b, where the transit radius yields an iron mass fraction of $45\%$, but when the oblateness is taken into account it reduces to $31\%$ (Dai et al, \emph{in prep}). Thus, some super-Mercuries may have more reasonable iron contents. However, this correction only applies to the closest planets, leaving several super-Mercuries in the sample population. Without correcting for detection bias, it is not possible to know how prevalent they are. But regardless of their prevalence, the existence of super-Mercuries demands an explanation. Here we discuss a few possibilities.

The leading theory for Mercury's large iron content involves a giant collision during formation that stripped its mantle \citep{Benz1988:Mercury, Benz08:Mercury}. However, investigating this scenario during formation with N-body codes has proven difficult to consistently form Mercury (similar location, mass and composition). Different works have increased the prevalence of Mercury formation from $<1\%$ to $\sim 5-10\%$ of the time by optimizing conditions that increase the likelihood of high-impact mantle-stripping collisions \citep{2021aClement, 2021bClement}, but one pervasive problem is the efficient re-accretion of debris onto the proto-Mercury that decreases the core-mass fraction (CMF) \citep{Scora2024}. 

In a more general sense, \citet{Scora2020} determined that for rocky exoplanets, the variation in core-mass fraction during formation only widens by $\sim 1.2$ due to giant impacts. This value increases slightly when considering high-eccentricity, high-inclination disks that produce more high-impact collisions, and assuming half of the debris is lost to radiation pressure \citep{Scora2022}. However, even when optimizing for mantle-stripping, very few planets increase their CMF by $\sim 1.5$x--corresponding to CMF=0.5 for the solar system \citep{Scora2022}. In addition, their results show that the variation in CMF decreases as planet mass increases. In other words, it is easier to form an iron-rich Mercury, than an iron-rich super-Mercury. This narrow variation at large masses is not enough on its own to explain some of the super-Mercuries in the sample of exoplanets (see Fig. \ref{fig:MR}). 

Even though collisions are expected to grow the terrestrial planets and perhaps rocky exoplanets, they may not explain the variety in the solar system or extrasolar systems. Another class of explanations for iron-rich compositions stems from mechanisms that separate metals and silicates in the solar nebula. Possibilities include the different temperatures at which metals and silicates condensate \citep{Lewis1972,Aguichine2020}, photophoresis arising from different conductivities and densities \citep{Wurm:2013}, differences in grain size owing to different surface tensions leading to different accreting scenarios \citep{Johansen:2022}. 

Here we explain two of these mechanisms. Photophoresis is a mechanism by which pebbles grow to planetesimals \citep{Cuello2016}, that also lead to a compositional sorting with iron rich planets forming in the inner regions of the disk \citep{Wurm:2013}. Photophoresis happens when gas particles in a low-pressure regime exchange momentum with a grain that has a temperature gradient established from being illuminated on one side by the star. The momentum exchange between the dust particle and the surrounding gas molecules leads to an overall force away from the radiation source resulting in a pile up of dust particles in ring-shaped distributions that favour subsequent clumping into larger pebbles. The dynamics of the grains depends on their density and conductivity resulting in a radial transport of silicate grains towards the outer regions of the disk that is more efficient than for iron-rich grains. With simple analytical arguments \citet{Wurm:2013} proposed this mechanism could explain Mercury and even super-Mercury as the sorting does not depend on the overall mass, and there is no shortage of material. However, this only works if the disk is sufficiently thin; by that time unsorted planetesimals may already had been formed.

Another possibility proposed uses nucleation theory in combination with the streaming instability mechanism. The streaming instability is a mechanism by which solids concentrate through aerodynamic interactions with the gas and can then become sufficiently packed to collapse under particle self-gravity, forming planetesimals. The threshold needed for triggering the streaming instability depends on the Stokes number, which determines the degree of coupling between the pebbles and the gas and which depends sensitively on the size of the particles. \citet{Johansen:2022} propose that because iron has a larger surface tension, it will nucleate around its own seeds and grow to larger sizes than silicates. This will lead to higher Stokes numbers that can trigger an accumulation into iron-rich planetesimals in the inner part of the disk \citep{Johansen:2022}.

\paragraph{Super-Moons} 

While many super-Earths fall within the refractory stellar compositions (see Fig. \ref{fig:MR}) there are also several that have larger radii, and as such are underdense. As mentioned earlier, these planets could have a volatile envelope above a rocky core, and the core composition could fit that of the star. The only way of knowing if they have an atmosphere is to use spectroscopy or phase curve observations. Given that phase curve observations may provide the minimum mass of the atmosphere, it may be possible to place constraints on the minimum amount of iron in these planets. Conversely, though, it is possible to know if these planets are bare rocks, in the sense that they do not have atmospheres affecting their radius (like the Earth and Venus). If so, they pose an interesting challenge: we have no reliable theories that can grow iron-poor silicate-rich planets. Collisions within an N-body framework have shown that in very special circumstances core-less planets can be formed but they are all small \citep{Scora2020, Scora2022}. A possibility, suggested by \citet{Dorn19:Ruby}, is that these planets have a substantial amount of Al and Ca and very small iron cores. They determind that abundances of 43\% of Al$_2$O$_3$ and 16\% of CaO in hand with no iron would explain the large radius of some low-density, low-mass planets. It is unclear if these abundances are feasible in astronomical contexts. As an example, Earth's content based on a pyrolitic model has Al$_2$O$_3$ at $4.45\%$ and CaO at a $3.55\%$ \citep{McDonough_Sun95}.

Two such intriguing underdense hot planets are Trappist-1b and c. These planets are the closest to the star of seven in the iconic Trappist-1 system. This system has special status because of how compact it is (with all planets inside 0.1 AU), and the different orbital resonances that allowed for exquisite mass data (errors of only 3-5\%) gathered through transit timing variations. Interestingly, the seven planets have compositions consistent with the same M-R relation, one that is less dense than Earth-like (iron mass fraction of 20\%, 
\citet{Agol:2021}). Their very compact configuration and orbital resonance chains point to a formation scenario where pebbles that originate in the outer disk accumulate at the H$_2$O ice line triggering planet formation via streaming instability, which may suggest some of these planets have water; these planets latter migrate to the inner disk in a process that results in an orbital resonance chain configuration \citep{Huang_Ormel:2022}. The inner planets have equilibrium temperatures that would yield vapor atmospheres susceptible to atmospheric loss, especially because M dwarf stars like TRAPPIST-1 have long active phases that can drive hydrodynamic escape for the first 1-2 billions of years \citep{Selsis2007}. 

While atmospheric features were attempted to be measured on all Trappist-1 planets through transmission spectroscopy with \textit{Spitzer} and HST, only constraints on the bulk metallicity were feasible -- the planets do not have hydrogen-rich atmospheres \citep[e.g.,][]{deWit2018, Zhang2018,Garcia2022}. JWST has the capability to probe with much better precision. Thermal emission data for the inner planets b \citep{Greene2023}, and c \citep{Zieba2023} is consistent with bare rocks without any CO$_2$ in their atmosphere, but the transmission spectroscopy observations continue to be plagued by degeneracies with stellar contamination \citep{Lim2023}. These findings point to the need of obtaining the composition of Trappist-1, the host star, to test whether an iron-depleted scenario can be the result of accreting around an iron-depleted star. Unfortunately, low mass stars have complicated molecular lines that make compositional inference extremely challenging. JWST observations of the outer planets have been taken and analysis is underway. Soon we will have a better understanding on which Trappist-1 planets have volatiles and which do not, setting constrains on how they formed. 

Observations of rocky exoplanets and their stars will help us understand better how major-element composition gets set during formation and will help put our Solar System into context. The jury is still out as to whether or not a primordial composition is valid for exoplanets, how prevalent are super-Mercuries and how they formed, and what is the nature and origin of the underdense planets. 

\subsection{Volatile Exoplanets: Mini-Neptunes, Hycean Worlds and Water Worlds} 
\label{volatileplanets}

Volatile exoplanets are characterized by being too large to be composed only of rocks. These are the planets that sit above the rocky threshold radius (RTR in Fig, \ref{fig:MR}). They must have a volatile component to explain their radius. The key problem becomes determining the composition of this layer, and what it says about its formation and/or evolution. 
Based on arguments of abundance, the most likely candidates are H, He and H$_2$O. Traditionally termed mini-Neptunes or sub-Neptunes, these volatile planets are now thought to have nuanced composition that has introduced new terminology. 

If planets have an atmosphere made of H-He, with or without water vapour they are considered ``mini-Neptunes". If they are cold enough to allow for water to be in condensable form at the surface below a H-He atmosphere, these are ``Hycean planets". If they only have water in condensed form mostly they are termed ``water worlds", but if these planets are too close to their star and instead have water in vapour form they are ``steam planets". The only family that is easy to label with Mass-Radius (M-R) data are the mini-Neptunes that are too large to be made of pure water vapour.\footnote{Notice though there are several pure water vapour lines according to different equilibrium temperatures so that comparisons to determine H/He based on MR only have to take into account the $T_{eq}$ of the planet.} The rest of planets require atmospheric composition characterisation to correctly identify them. Within the radius range of $1-2R_E$ these measurements are extremely difficult. However, JWST is providing important observations in this direction, including the atmospheric characterisation of K2-18b which according to thermochemical interpretations of its spectra, appears to be a strong candidate for a Hycean world \citep{Madhu23}. 

\paragraph{Primordial Atmospheres}
The discovery of mini-Neptunes at short and intermediate periods poses a problem to the standard theory of solar system formation (mentioned in section \ref{formation}) that in broad terms predicted a clear threshold between small bare planets and large gaseous ones. \citet{Pollack1996} predicted that planets below a threshold of $\sim 10 M_E$ would acquire no gas; planets that reached a mass of $\sim 20 M_E$ split equally between gas and rocks would undergo runaway accretion and become gas giants; and planets in between would acquire a smaller gaseous atmosphere. This theory largely explained the Solar System, including Uranus and Neptune. Being at large semi-major axes, these ice giants took such a long time to form that by the time they reached a considerable size the solar nebula (the sun's protoplanetary disk) had already dissipated, leaving them without the ability to accrete large atmospheres. \citet{Rafikov06} showed that this threshold in mass was dependent on the disk location, opacities, accretion luminosity and mean molecular weight so that some overlap between 5-20 $M_E$ could be expected within the Terrestrial region. 

The advent of volatile planets showed instead that within the same mass-radius space of $1-10 M_E$ and $ 1-2 R_E$, planets with and without small envelopes would co-exist (see Fig. \ref{fig:MR}), defying the original simple formation theory. Since then, there have been numerous studies that have aimed to explain the formation of mini-Neptunes conceived as planets that had retained small primordial atmospheres of H/He. A few ideas put forward involve invoking a gas-poor environment due to a timing issue: either because the gas disk lifetime is short \citep{Alibert06, Terquem07} or the planets formed late when there was little gas around \citep{Lee_Chiang16, Dawson16}. Alternatively, small planets (with cores of $<15 M_E$) may accrete gas more slowly than expected due to 3D flow in the envelope that slows cooling and hence gas accretion \citep{Artymowicz1987,Wu2013,Ormel2015,Lambrechts_Lega17}. 
If the planet indeed retained these type of atmospheres, the expected composition would be hydrogen and helium in or close to stellar relative abundances, or with some helium enhancement due to preferential escape from H \citep{Malsky2023}. These are relatively easy atmospheres to detect owing to their large scale height. 

A second possibility for a primordial atmosphere is the consideration of a steam planet where water was acquired in condensed form under cold conditions and subsequently evaporated at the surface. This can happen in two ways. The planet could form beyond the snow line and later migrated inwards to hotter regions \citep{Burn2024}. Or, the planet could form in the hot inner region and have water delivered by an influx of cold wet planetesimals from the outer part of the disk, similar to the Terrestrial planets. The expected amount of water would be different depending on the location of formation (inside versus outside the iceline) and accretionary pathway. This water is expected to have traces of methane and ammonia as these also condense in the outer portions of the disk.

Within this context, a Hycean world is a combination of both. A planet that acquired some H and He, along with water, and that is cold enough for the water to have condensed out into an ocean. K2-18b is a prime candidate because it has an extended atmosphere [1000 km above a pure rock and pure water line, \citet{Madhu20}], that requires some H/He. With an equilibrium temperature of only $\sim$250K, it may have a cool atmosphere that has kept an ocean. While initial observations with the Hubble Space Telescope had hinted at the possibility of water with a signature near 1.4 $\mu$m \citep{Tsiaras:K2-18b,Benneke:K2-18b}, recent JWST observations that span a much wider frequency range (1 $\mu$m -5 $\mu$m) ruled water out in favor of CH$_4$, and detected the presence of CO$_2$ (Fig. \ref{fig:K2-18b}). Both species at the same time would suggest there is enough hydrogen and oxygen to have water, so that the lack of a detection is suggestive this water is not in gaseous form, but instead condensed either in clouds and/or an ocean underneath \citep{Madhu23}. Alternative, it has been suggested that the JWST detections could be explained with a gas-rich mini-Neptune with 100 $\times$ solar metallicity, where CH$_4$ and CO$_2$ are produced thermochemically in the deep atmosphere \citep{Wogan2024}. A yet different scenario argues that the planet might be too hot to host liquid water and that the presence of CH$_4$ and CO$_2$ and the absence of NH$_3$ observed by JWST could result from the interaction of a thick hydrogen envelope with a molten planetary surface \citep{Shorttle2024}. Observations of the CO$_2$ and CO spectral features $>$ 4 $\mu$m can help distinguish between the three scenarios (ocean world vs. mini-Neptune vs. magma; \citeauthor{Shorttle2024} 2024).  

\paragraph{Secondary Atmospheres}
\label{secondaryatmospheres}

The other possibility for volatile planets is that their atmospheres are of secondary origin and outgassed\footnote{The process of outgassing involves the release of gases from melt. For some planets, temperatures may be so high that miscibility may prevent a clean separation of liquid/solid surface from a gaseous envelope} from the interior. 
In this section we are mostly interested in extended atmospheres that can affect the radius in a measurable way. Planets like Earth or even Venus have atmospheres too thin to be reflected in an M-R diagram. 

One possibility for these secondary extended atmospheres is hydrogen that was stored within the magma ocean as it was ingassed from the stellar nebula, or delivered by impacts, and subsequently outgassed. \citet{Chachan_Stevenson18} time-evolution calculations considered the magma ocean as a reservoir for hydrogen that replenishes an escaping atmosphere. Based on solubility laws that dictate how much H is in the atmosphere versus magma ocean, they find that planets end up with a $\sim 0.1\%$ H envelope, which can explain some of the low-density low-mass exoplanets. This is a different way to get a low-molecular envelope that involves reprocessing from the interior. While both scenarios of a primordial versus an outgassed origin for the envelope can lead to H-dominated envelopes, the relative abundances of other compounds should differ owing to different solubilities in the magma. 

Another possibility is a water atmosphere in contact with a magma ocean. Based on solubility laws as well, \citet{Dorn:2021} determined that planet with 5\% by weight of water have a vapour atmosphere extending $\sim 1300$ km beyond a wet magma ocean. This water envelope could have been outgassed from the magma ocean. Similarly, \citealt{Kite2016} suggested that endogenic water in the atmosphere, that is water that was created from the chemical interaction of nebular hydrogen (H) with the liquid iron oxide (FeO) in the magma ocean, could explain the radii of planets at intermediate periods (10-100 days) just below the radius valley. At shorter periods, relevant to USPs, the hydrogen and water would be lost due to atmospheric evaporation. In contrast, recent solubility calculations in by \citep{Sossi:2023} have suggested that most of the water would be in magma ocean severely limiting the size of steam atmospheres.

A third possibility invokes the presence of an atmosphere in contact with a magma ocean as well, but dominated mostly by carbon, either in the form of CO or CO$_2$, depending on the redox state of the planet. \citet{Peng2024} considered a C-O-H atmosphere with C contents corresponding to chondritic abundances ($\times \sim 3$ higher than Earth's bulk budget) and obtained that C dominated atmosphere would extend the planetary radius by 1000-1500 km, and explain some low-density, highly irradiated planets such as TOI-561b, and 55 Cnc-e. 

Exactly what compounds are outgassed from the Earth from a magma ocean will depend on the chemical budget (the amount of C, H, O, and other compounds) and the redox state (\citet{Gaillard:2022, Bower:2022, Charnoz:2023, Peng2024} ). It is difficult to predict a-priori the redox state of a system, but perhaps atmospheric characterisation can help us piece together the environment under which outgassing proceeded.

\subsection{Ultra Short Period Planets - Lava Planets}
\label{USP}

\begin{figure}
     \centering     \includegraphics[scale=0.90]{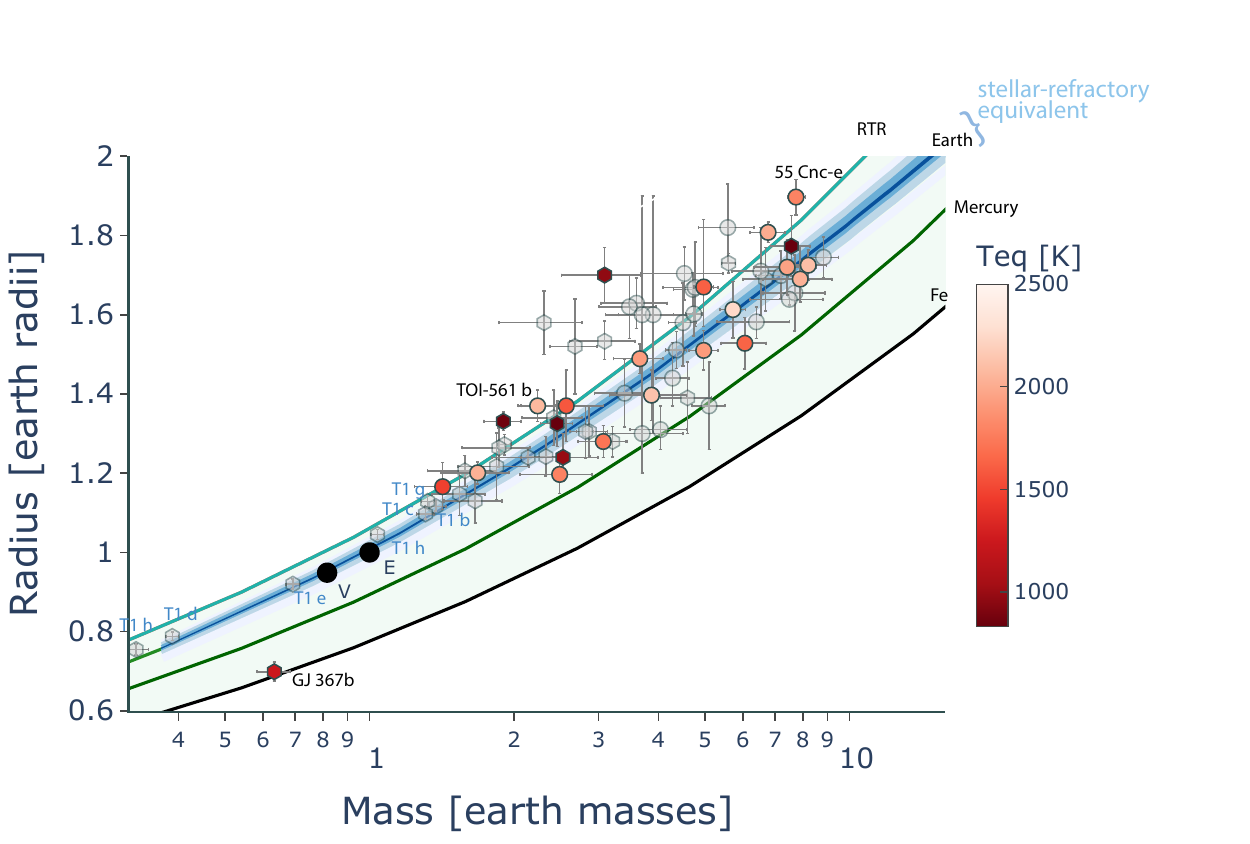}
     \caption{Ultra Short Period Planets with mass and radius errors lower than 20\% color coded according to their equilibrium temperatures calculated with a Bond albedo of 0.3 and redistribution factor of 4. Planets around FGK stars are shown with circles, and planets around M stars are shown with hexagons. Grey planets have periods of more than one day. For reference solar system planets are shown as black symbols. The shaded blue lines correspond to the M-R relations of stellar refractory ratios.}
     \label{fig:USP}
 \end{figure}

The last type of planet that we will discuss are the so-called ``ultra short period" (USP) planets. These compact planets (with $R< 2R_E$) have periods of less than one day, and given their proximity to their parent star, they are presumed to be tidally locked, showing the same side to the star during their orbit. Given their M-R data (see Figure \ref{fig:USP}) and their warm/hot surfaces also, extended H or H$_2$O atmospheres can be ruled out, making these planets primarily rocky with no degeneracy. 

The compositional spread of the USP planets with a measured density is lower than for the full population of rocky exoplanets, with many that orbit around FGK stars having M-R data consistent with an Earth-like composition \citep{Dai2021}. However, very intriguingly, some of them are underdense, including planets that require a volatile envelope to explain their large radius (those that are above the RTR in Figure \ref{fig:MR}).

For those that have an atmosphere, it would be highly vulnerable to loss owing to the high irradiation from the star. However, for those around FGK stars their equilibrium temperatures are so hot ($T_{eq} > 1400$K) that these USP most likely have a surface temperature above that of molten rocks [$> 1500 K$, \citet{Hirschmann2020}] such that these planets have surface magma oceans that could replenish the atmosphere with volatiles. Those around M dwarf stars may not be irradiated enough (with $800 \mathrm{K} < T_{eq} < 1400$K) to sustain magma oceans underneath \citep{Selsis2023}. Thus, the scenarios of ``Steam Worlds" \citep{Kite2021, Dorn:2021} and ``Puffy Venuses" \citep{Peng2024} may apply to these USPs, and lend themselves as unique windows to understanding atmosphere-mantle connections, applicable to cooler planets as well. They may be a window into the early epochs of Earth and/or Venus. Only through observations, will it be possible to determine if they indeed have an atmosphere. Interestingly though, should some of them be bare rocks, phase curves/secondary eclipse observations would be probing the surface directly. Based on emissivity of different minerals, these observations may constrain the composition of their surfaces directly \citep{Hu:2012:surfaceemissivity}.  

Owing to their brightness, some of the USPs are amenable to phase curve/secondary eclipse observations. Indeed, already there is a growing list of USPs that have been allocated JWST observational time.  
Here we discuss a few features of these planets.

The most studied USP is 55 Cnc-e. Different mass and radius measurements suggest either the planet requires volatiles \citep{Demory2011,Winn2011,Demory2016,Crida2018,Dai2019}, or one that intersects the rocky region \citep{Bourrier2018}. Its mass and radius are close to $8M_E$ and $1.9 R_E$. Phase curve data with Spitzer \citep{Demory2016} showed an offset of 41 degrees indicating an atmosphere redistributing heat between the night and the dayside, but a high difference in temperature of 1300K indicative of an atmosphere that does not globally carry heat. Their conclusions were that either the planet has an optically thick atmosphere redistributing heat only on the dayside or a planet with lava currents carrying heat from the day to the night side. \citet{Angelo2017} reanalyzed the data and ruled out the lava scenario based on timescale arguments. Their model suggested a similar timescale for radiative and advective flow which they claimed could not be attained by magma lake flows based on models by \citep{Kite2016}. Later reanalysis of the \textit{Spitzer} data by \citet{Mercier2022} suggested a hot-spot offset value consistent with zero and a day-night temperature difference of 2700 K, which are both parameters consistent with a no-atmosphere scenario. Ground and space observations have looked at signatures for evidence of atomic or molecular lines, but had yielded non-detections of a large list of compounds H$_2$O, TiO \citep{Esteves2017, Jindal2020}, CO, CO$_2$, HCN, NH$_3$, and C$_2$H$_2$ \citep{Deibert2021}, as well as Mg, Fe, Si, O, Na, Ca, Al, Ti P, F, S and others \citep{Keles2022}. There is only one tenuous suggestion of Na from the D lines (at a 3$\sigma$ level) and ionized Ca from the H\&K lines [seen in only one transit, \citet{Ridden-Harper2016}] but given the weak signal the authors did not claim detection. 

Fortunately, this planet has recently been observed with JWST. \citet{Hu2024} obtained thermal emission spectrum between 4 $\mu$m to 12 $\mu$m and observe an intriguing feature at 4.5 $\mu$m. They obtain a brightness temperature of $\sim 1800$K, lower than the temperature expected at zero heat distribution with zero albedo ($\sim2500$ K), strongly ruling out the possibility of a lava world shrouded by a vaporized rock atmosphere. Furthermore, they claim their spectrum shows a feature at 4-5 $\mu$m that is best described by absorption by CO$_2$ or CO. A secondary atmosphere may also explain the variability seen in the 4.5 $\mu$m feature seen in earlier data from \textit{Spitzer} \citep{Demory2016,Tamburo2018} from changes in atmospheric composition coming from the exchange with the magma ocean (see Fig. \ref{fig:55Cnc-e}).

We anticipate more observations of 55 Cnc-e in the future with JWST given it is high-value: it defies conventional formation theories with its underdensity, has atmospheric measurements that seem in tension, and its extreme nature allows us to look into our past when the Earth had magma oceans. 

Very interestingly, there is another ultra hot under dense planet, TOI-561b that will soon be observed by JWST \citep{Teske2023}. TOI-561b shares similar characteristics as 55 Cnc-e being close to the boundary between definitely volatile and possibly rocky, but much smaller at $2.24 M_E$ and $1.37 R_E$ \citep{Brinkman2023}. Testing whether its atmosphere/surface resembles 55 Cnc-e or not is of utmost value.

On the other side of the USPs is GJ~367b with a higher density than Earth's. 
Based on revised mass values and with corrections its radius from tidal distortion (Lee et al., in prep), we estimate core-mass fraction values of 58-74\% (based on models by \citet{Plotnykov20}).
Recent JWST observations by \citet{Zhang2024} have provided thermal emission consistent with an airless body with no heat redistribution and a low albedo of $\sim$0.1 (similar to that of Mercury and the Moon). The dayside temperature is close to $\sim$1400K and suggestive of at least a localized magma ocean that could sustain an atmosphere. However, the lack of redistribution of heat places constraints on an atmosphere of at most 0.1 bars, which is also susceptible to evaporation. Thus, \citet{Zhang2024} concluded that this planet was either formed dry or desiccated by evaporation. 

These are just the first results on the characterization of small planets' atmospheres. Even just determining which ones have atmospheres is valuable, given the extreme conditions where exoplanets are found. 

\section{Polluted White Dwarfs: A Unique Window into Rocky Planet Compositions}
\label{WhiteDwarfs}

As discussed in section \ref{formation}, there is ample evidence for disks of second-generation debris around stars of many ages, well into mature systems including our own Solar System. And yet it was somewhat curious when infrared excess was observed around a cool white dwarf star \citep{Zuckerman1987}, and perhaps even more surprising when metals were detected in its atmosphere \citep{Koester1997, Zuckerman1998}. Given the extremely high gravities of these remnants of low-mass main sequence stars, white dwarf atmospheres are supposed to have pristine compositions consisting of H (DA white dwarfs) or He (DB white dwarfs) only. This is because any heavier elements would have sunk on a timescale of 10$^{-2}$--10$^{7}$ years due to radiation pressure no longer being strong enough to keep them aloft as the white dwarf cools \citep{Chayer1995}. But in a subset of white dwarfs (now estimated to be 25-50\%), many of which also have detected debris disks, metals including Ca, Mg, Si, Fe, Ni, and even more volatile elements like C, N, and O have been detected \citep{Klein2010, Gansicke2012, Jura2014, Xu2014, Xu2019, Bonsor2020}.

In a pivotal paper, \cite{Jura2003} showed that these observations could be matched by a minor planet around a white dwarf being tidally disrupted, producing a cascade of collisions that grind the solids into dust that forms a disk and gradually accretes onto the white dwarf surface, polluting its atmosphere. Subsequent studies estimated the disrupted solid bodies could range from 100 km to the size of asteroids and dwarf planets \citep{Zuckerman2010}. The resulting dust disks were readily detected by \textit{Spitzer} as infrared emission in excess of that expected from the hot white dwarfs. The latter, due to their very high temperature, emit mostly in the UV and the optical. These dust disks are inferred to be very compact, $\sim$ 1~$R_{\odot}$ in size \citep{Jura2007, Jura2009,Farihi2016,Wilson2019}. Some also show evidence of circumstellar gas seen in emission or absorption, most likely released from the tidal disruption of the planetesimals whose fragments are exposed to the intense radiation from the white dwarf, leading to the sublimation of their volatile component (a process known as sublimative erosion) \citep{Gansicke2006b, Gansicke2012, Debes2012}. 

Extensive theoretical investigations have been carried out to explore the origin of the bodies contributing to the debris dust and the atmospheric pollution (for a thorough discussion see \citeauthor{Veras2021}, 2021). These investigations generally evolve the planetary system and the star simultaneously. This is because the phases previous to the formation of the white dwarf involve significant stellar mass loss that can affect the dynamics of the planetary system and the planetary architecture, and changes in stellar luminosity and temperature, that can impact the planets and planetesimals closest to the dying star. These investigations explore the contribution of different populations of minor bodies analogous to those in the asteroid, Kuiper and scattered belts, the Oort cloud, and the moons, which will contribute differently to the mineralogy and elemental abundances of the white dwarf debris and atmospheric pollutants. The models also explore outcomes in the absence and in the presence of planets, for single planet system and multiplanet systems. The presence of one or more planets aids the delivery of planetesimal(s) into the tidal radius of the white dwarf, with multiplanet systems able to trigger multiple planetesimal deliveries resulting from separate dynamical instability events \citep{Veras2021}. The physical processes considered by these studies include the effect of gravitational interactions, stellar flybys and the Galactic tides (the latter two affecting planetesimals in the outermost system), and rotational fission due to spin angular momentum exchange and sublimation (for planetesimals closer in), among others \citep{Veras2021}. 

Almost all exoplanets known so far orbit stars that will eventually become white dwarfs. We now know that the occurrence rate of planetary systems around white dwarfs and main-sequence stars is comparable and that planetary systems in different stages of destruction have been observed around white dwarfs (from intact planets to small planets and planetesimals shredded to their elemental constituents) \citep{Veras2021}. These include the detection of at least one and probably several disintegrating planetesimals around four white dwarf systems \citep{Vanderburg2015, Manser2019, Vanderburg2020, Guidry2021}, and the discovery of giant planet candidates around another four white dwarfs \citep{Vanderburg2020, Luhman2011, Gansicke2019, Thorsett1993, Sigurdsson2003}, in addition to the detection of more than 40 white dwarf debris disks and of more than 1000 white dwarfs with signs of photospheric pollution \citep{Farihi2016, Veras2021}. 

The atmospheric abundances of polluted white dwarfs (PWDs) can constrain to an extraordinary degree the bulk elemental composition of the ``left over'' building blocks of planets, providing a more direct perspective on the compositions of exoplanets than is possible via observations of fully-formed, mature exoplanets and/or their host star abundances. Notably, through mostly UV (HST) studies, thus far most of these PWDs show compositions similar to the bulk Earth in that they are dominated by O, Fe, Si, and Mg \citep{Xu2019}. There are a few exceptions where the accreted material seems to be water-rich \citep{Farihi2013, Raddi2015, GentileFusillo2017}, analogous to a Kuiper-Belt Object (enhanced oxygen, carbon, and nitrogen; \citealt{Xu2017}), or even resembling giant planets \citep{Gansicke2019} or icy moons \citep{Doyle2021}. But the overall trends on PWD compositions suggest that Solar System-like geologies might be common \citep{Jura2014,Harrison2018}, even more specifically enhanced oxidizing conditions for rocky body formation, versus the more reduced primordial solar composition \citep{Doyle2020}. Recently, studies have explored in more detail the mineralogies implied by PWDs compositions, based on siderophile versus lithophile elements, 
with some identified as ``core''- or ``crust''- or ``mantle''-like \citep{Hollands2018, Hollands2021, Putirka2021, Bonsor2023}. Some authors have suggested the relative amounts of these different types of material could even be used to infer core-mantle differentiation and thus the frequency of heating by short-lived radioisotopes \citep{Jura2014,Curry2022}.

However, the interpretation of PWD abundances in these ways is challenging due to their small overall numbers as well as large errors; \cite{Trierweiler2023} recently showed that the majority of PWDs are indistinguishable from chondritic compositions, versus crust, and that determining detailed mineralogies (at least distinguishing between olivine, orthopyroxene, and clinopyroxene) is not possible given current uncertainties. \cite{Doyle2020} also pointed out that more reduced bodies polluting WDs are harder to detect and accurately measure with current methods. Another challenge to the interpretation of individual observations is the assumption that the the different components of the accreted planetesimal (core/mantle, volatile/refractory) are always incorporated into the white dwarf in similar proportions. It has been proposed that the accretion might be done asynchronously and that the volatile component that sublimates and is not blown out by radiation pressure, can accrete onto the white dwarf before the refractory component \citep{Malamud2016, Brouwers2023}, in which case the polluting signature may not be always representative of the planetesimal bulk content. 

The superb sensitivity and spectroscopic capabilities of JWST have recently opened a window to study the spatial distribution of the debris dust around the white dwarf, probing the wider system architecture, and the composition of the debris dust, but this time in terms of its mineralogy rather than the elemental abundances inferred from the white dwarf atmospheric pollution. Indeed, the first JWST observations of a white dwarf debris disk show 9–12$\mu$m solid-state emission features similar to the silicate minerals observed in debris discs around main sequence stars, and a tentative detection of 7$\mu$m emission features that could be associated to carbonates and could indicate aqueous alteration in the disrupted parent body \citep{Swan2024}. There is a promising future for comparative studies of polluted white dwarfs, with thousands of candidates identified by \textit{Gaia} \citep{GentileFusillo2019} that will provide fruitful for follow-up to try to detect disks and metals in their atmospheres.

\section{Conclusion}

The field of exoplanet research has made significant strides over the last 30 years. We are finally entering the era of super-Earth and mini-Neptune characterization, with better mass and radius measurements that provide tighter constraints on bulk composition and basic information on the atmosphere. Studying diverse exoplanets from as many different angles as possible, from formation theories, bulk chemistry and atmospheric characterisation, to the final fate of being devoured by the central stellar, will enable us to understand their origins under a general framework, allowing us to piece together a broader view of the formation of our own solar system.

Focused on low-mass exoplanets, we saw they exhibit great compositional diversity. Some are compact, similar to Mercury, many seem to be like Earth, while others seem to be iron-poor yet still rocky. Within these type of planets, a fundamental first order question is to know whether or not rocky planets share the composition of their parent star in terms of refractory ratios pointing to a primordial, unprocessed origin. 
Among these rocky planets, there is a growing list, including planets more massive than Earth, that seem consistent with an airless body, much like Mercury and Mars, and less like Venus or Earth. This list will continue to grow and build our understanding of which planets build and retain atmospheres. Among large planets that cannot be rocky, some require H/He, while others may have water atmospheres. New results from JWST on the volatile-rich planet K2-18b are consistent with a Hycean world (with a H/He atmosphere over liquid water), a mini-Neptune or a magma-ocean planet. By knowing the atmospheric chemical ratios of planets the hope is to distinguish between the different possible scenarios and to piece together the origin of the atmosphere, whether primordial or of secondary nature. 

A subset of exoplanets are exposed to such extreme irradiation from their star that their surface should melt, whether directly or beneath an atmosphere if present. As a class, these ultrashort period planets lend themselves to understanding chemical exchange between a magma ocean and an atmosphere, or directly probing an exoplanet surface. One iconic case, 55 Cnc-e, was recently observed by JWST revealing that it has a carbon-dominated atmosphere. Straddling the boundary between rocky and volatile, this planet, as others sharing its characteristics, defy conventional theories on atmospheric escape and planet formation. 

As a complementary way to understand the composition of exoplanets, observations of polluted white dwarfs are delivering intriguing results. Many of these stars are showing that the material they swallowed has geochemical signatures not very different from the inner solar system, and while the precision may not be enough to distinguish between different silicate rock mineralogies, it is enough to infer iron differentiation in exoplanets. On a different angle, the study of protoplanetary and debris disks sheds light on the formation and evolution of planetary systems; from the search for volatiles in disks we can learn about the volatile budget of planets forming at those disk locations and the possible source of the volatiles; the long-term monitoring of dusty debris disks can inform us of the role of giant collisions in planetary system formation and evolution. Lastly, exoplanets afford us the possibility to probe how much Jupiter and the giant planets may have shaped the solar system formation, and perhaps the habitability of Earth, as exoplanet systems exhibit a diversity of configurations with and without giant planets.
  
Looking ahead, we can expect a wealth of data from JWST observations and the different space and ground facilities that will be upgraded, have been approved, or are being planned. These facilities include the approved space missions PLATO (2026), Roman (2027), and Ariel (2029); the Habitable World Observatory (late 2030's-early 2040's), recommended by the National Academies’ Pathways to Discovery in Astronomy and Astrophysics for the 2020s; and ground-based facilities like future upgrades to the ALMA, a newly proposed single-dish Atacama Large Aperture Submillimeter Telescope, the Next Generation Very Large Array, and extremely large ($\geq 25$ m aperture) ground-based telescopes (Extremely Large Telescope, the Giant Magellan Telescope, the Thirty Meter Telescope). The future is bright for exoplanets.

\section*{Acknowledgements}
We would like to thank Prof. D. Fisher for her insightful comments as a reviewer, and Dr. A. Morbidelli for enlightening discussions and valuable feedback on the manuscript. DV wishes to acknowledge funding from the Natural Sciences and Engineering Research Council of Canada (Grant RGPIN-2021-02706).

\bibliographystyle{apalike}
\bibliography{references_geochemistry}

\end{document}